\documentclass[twocolumn,aps,prd,groupedaddress,nofootinbib,superscriptaddress,floatfix,amsmath,amssymb,preprintnumbers]{revtex4-1}
\usepackage{graphicx}
\usepackage{dcolumn}
\usepackage{bm}
\usepackage{slashed}
\usepackage{siunitx}
\usepackage{hyperref}
\usepackage[mathlines]{lineno}

\usepackage{comment}
\usepackage{soul}
\usepackage{xcolor}
\hypersetup{colorlinks, linkcolor = [rgb]{0,0.0,0.5}, citecolor = [rgb]{0,0.0,0.5}, urlcolor = [rgb]{0,0.0,0.5}}

\allowdisplaybreaks[1]

\begin{document}


\title{Power corrections in semi-inclusive deep inelastic scatterings at fixed target energies}

\newcommand*{\JLAB}{Theory Center, Thomas Jefferson National Accelerator Facility, Newport News, VA 23606, USA}\affiliation{\JLAB}
\newcommand*{\DUKE}{Department of Physics, Duke University, Durham, North Carolina 27708, USA}\affiliation{\DUKE}

\author{Tianbo~Liu}\affiliation{\JLAB}\affiliation{\DUKE}
\author{Jian-Wei~Qiu}\affiliation{\JLAB}


\begin{abstract}
The COMPASS collaboration published precise data on production cross section of charged hadrons in lepton-hadron semi-inclusive deep inelastic scattering, showing almost an order of magnitude larger than next-to-leading order QCD calculations when $P_{h_T}$ and $z_h$ are sufficiently large.
We explore the role of power corrections to the theoretical calculations, and quantitatively demonstrate that the power corrections are extremely important for these data when the final-state multiplicity is low and the production kinematics is near the edge of phase space. Our finding motivates more detailed studies on power corrections for upcoming experiments at Jefferson Lab, as well as the future Electron-Ion Collider.
\end{abstract}

\maketitle

\section{Introduction \label{sec:intro}}

Unveiling the structure of nucleons (or hadrons, in general) in terms of quarks and gluons of quantum chromodynamics (QCD) is one of the central goals that has been actively pursued by the science community since the first lepton-proton deep inelastic scattering (DIS) experiment took place at SLAC about 50 years ago~\cite{Bloom:1969kc}. However, owing to the fact that no isolated quarks and gluons have ever been seen in a modern detector, nearly all analyses of high energy scattering events with identified hadron(s) rely on the QCD factorization theorem~\cite{Collins:1989gx} that provides the link between the observed hadrons and the quarks and gluons, or collectively, partons, that participated in hard scatterings.  The inclusive lepton-proton DIS at a large momentum transfer, $Q\gg \Lambda_{\rm QCD}\sim 1/\text{fm}$, is dominated by the scattering of the lepton off one active quark/parton inside the colliding proton~\cite{Feynman:1969ej}.  With one hard momentum scale $Q$, the inclusive DIS cross section is not very sensitive to the dynamics at a typical hadronic scale $ \sim 1/\text{fm} \ll Q$, and can be factorized into the lepton-quark scattering at a short-distance ($\sim 1/Q$) multiplied by corresponding quark parton distribution functions (PDFs), $\phi_{i/P}(x,\mu^2)$, interpreted as the probability distribution to find this active quark of flavor $i$ inside the colliding proton at the factorization or probing scale $\mu \sim Q$, carrying the proton's momentum fraction $x$.  This factorization is known as the QCD collinear factorization, with active quark's transverse momentum $k_T$ integrated into the PDFs and overall corrections suppressed by inverse powers of $Q$.  The measurement of the inclusive DIS cross section has provided good information on the proton's partonic structure, encoded in these factorized PDFs.  

It is the QCD factorization that provides the ``probe'' --the short-distance partonic scattering to enable us to ``see'' quark, gluon and their dynamics indirectly.  The predictive power of such factorization approach relies on both the {\it precision} of the probe, which we could achieve and improve by calculating the partonic interactions at the scale $Q$ order-by-order in QCD perturbation theory, and the {\it universality} of these PDFs, so that we are able to extract them from data of some experiments and use them to predict and to be tested in other measurements.  In terms of this factorization formalism, QCD has been extremely successful in interpreting almost all available data from high energy scatterings with probing distance less than 0.1\,fm (or equivalently, with the momentum transfer greater than 2\,GeV)~\cite{Brambilla:2014jmp,Gao:2017yyd}.  It is this success that has provided us the confidence and the tools to discover the Higgs particle and to explore new physics beyond the Standard Model of particle physics in high energy hadronic collisions~ \cite{Cepeda:2019klc,Azzi:2019yne,CidVidal:2018eel}.

\begin{figure}[htp]
\centering
\includegraphics[width=0.18\textwidth]{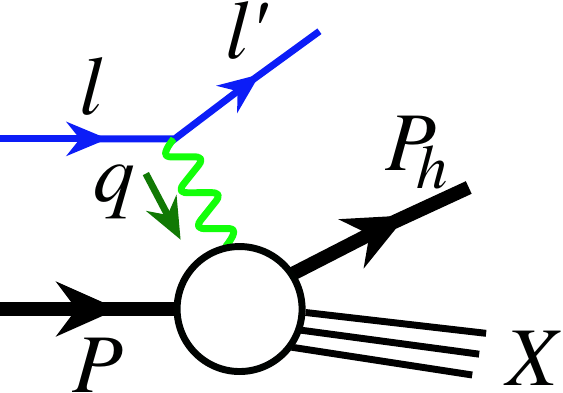}
\caption{Sketch for lepton-hadron semi-inclusive DIS.}
\label{fig:sidis}
\end{figure}

Instead of summing over all hadronic final states, the semi-inclusive DIS (SIDIS) identifies one final-state hadron of momentum $P_h$, as illustrated in Fig.~\ref{fig:sidis}, and covers a part of the inclusive lepton-hadron DIS cross section.     
By detecting one hadron in the final-state, SIDIS enables us to explore the emergence of color neutral hadrons from colored quarks and gluons, in addition to the information on finding a quark or gluon inside the colliding hadron.  By selecting different type of observed hadrons (pion, kaon, ...), SIDIS provides opportunities to study the flavor dependence of QCD dynamics.  With a large momentum transfer carried by the virtual gauge boson of momentum $q$ ($Q^2 \equiv -q^2 \gg 1/\text{fm}^2$), as shown in Fig.~\ref{fig:sidis}, SIDIS could provide a short-distance probe with an additional and adjustable momentum scale by measuring the hadron at different momentum $P_h$.  In the frame where the exchange virtual gauge boson of momentum $q$ and the colliding hadron of momentum $P$ are headed on, the leading contribution to the SIDIS is naturally from the region where the transverse momentum of the observed hadron, $P_{h_T}\ll Q$, and the scattering provides a short-distance probe with two very different momentum scales, from which the harder scale $Q$ localizes the hard collision to ``see'' the particle nature of quarks and gluons, while the soft scale $P_{h_T}$ is sensitive to the confined motion of quarks and gluons in the direction perpendicular to the direction of the colliding proton.  In this kinematic regime where $Q^2\gg P_{h_T}^2\gtrsim 1/\text{fm}^2$, similar to the inclusive DIS, SIDIS cross section can be factorized into a product of perturbatively calculable lepton-parton scattering at the hard scale $Q$, corresponding transverse momentum dependent (TMD) parton distribution functions (or simply, TMDs), $\phi_{i/P}(x,k_T,\mu^2)$ with $k_T$ being the active parton's transverse momentum perpendicular to the direction of the colliding hadron of momentum $P$, and TMD fragmentation functions (FFs), $D_{j\to h}(z,p_T,\mu^2)$ with the emergent hadron-type $h$ carrying momentum fraction between $z$ and $z+dz$ of the fragmenting parton of momentum $p$ and $p_T$ being the parton's transverse momentum off the direction of the observed final-state hadron of momentum $P_h$, where $i,j=\{q,\bar{q},g\}$ represent the active parton flavors~\cite{Ji:2004wu,Collins:2011zzd}.  In terms of this TMD factorization formalism, with the corrections of ${\cal O}(P_{h_T}/Q)$, lepton-hadron SIDIS is an excellent process to probe three-dimensional (3D) confined motion of quarks and gluons inside a bound proton, and has been actively pursued by experimental programs at all lepton-hadron scattering facilities, such as COMPASS at CERN \cite{Aghasyan:2017ctw} and various experiments at Jefferson Lab \cite{Dudek:2012vr}, as well as the future Electron-Ion Collider (EIC) \cite{Accardi:2012qut}.

On the other hand, when $P_{h_T}\sim Q\gg \Lambda_{\rm QCD}$, the SIDIS is dominated by a single hard scale, no longer sensitive to the dynamics at the scale of active parton's transverse momentum $k_T$ (or $p_T$), which is typically much less than $Q$, and the cross section should be better described by the QCD collinear factorization approach.  At the leading power (LP) of momentum transfer, the collinearly factorized SIDIS cross section is given by 
\begin{widetext}
\begin{eqnarray}
\frac{d\sigma_{l+P\to l'+P_h+X}}{d^3{\bf l}^{'}{\hskip -0.04in}/{\hskip -0.01in}(2E')\, d^3{\bf P}_h{\hskip -0.015in}/{\hskip -0.01in}(2E_h)}
\approx 
\sum_{i,j}\int_{x_B}^1 \frac{dx}{x} \int_{z_h}^1\frac{dz}{z^2}\, \phi_{i/P}(x)\, D_{j\to h}(z)\, 
\frac{d\hat{\sigma}_{l+i\to l'+j+X}}{d^3{\bf l}^{'}{\hskip -0.04in}/{\hskip -0.01in}(2E')\, d^3{\bf p}{\hskip -0.015in}/{\hskip -0.01in}(2E_p)} \, ,
\label{eq:fac-sidis}
\end{eqnarray}
\end{widetext}
where Bjorken variables are
\begin{equation}
x_B=\frac{Q^2}{2P\cdot q}, 
\quad
z_h=\frac{P\cdot P_h}{P\cdot q}, 
\label{eq:xbzh}
\end{equation}
and $d\hat{\sigma}_{l+i\to l'+j+X}$ is the perturbatively calculable partonic hard part for the lepton to scatter off a collinear on-shell parton of momentum $xP$ and flavor $i$ to produce an active parton of momentum $p=P_h/z$ and flavor $j$, which fragments into the observed hadron of momentum $P_h$.  The $\phi_{i/P}(x)$ and $D_{j\to h}(z)$ in Eq.~\eqref{eq:fac-sidis} are collinear PDFs and FFs, respectively, and their factorization scale dependence is suppressed.  The corrections to Eq.~\eqref{eq:fac-sidis} are suppressed by $1/P_{h_T}^2$ (or $1/Q^2$) for spin-averaged cross sections.  A smooth transition and matching between the TMD and collinear factorization approaches has been developed and tested for various two-scale observables, {\it e.g.},  heavy gauge boson $W/Z$ production in hadron-hadron collisions~\cite{Collins:1984kg,Qiu:2000hf,Landry:2002ix}. 
However, the precise SIDIS data, recently published by COMPASS Collaboration~\cite{Aghasyan:2017ctw}, show almost one order of magnitude discrepancy between the data and the leading order (LO) theoretical prediction for the region where $P_{h_T}\sim Q$ and $z_h$ are large~\cite{Gonzalez-Hernandez:2018ipj}, and the next-to-leading order (NLO) corrections provide very little help to reduce this discrepancy~\cite{Gonzalez-Hernandez:2018ipj,Wang:2019bvb}.  Soft gluon resummation does help enhance the prediction of theoretical calculations for the high $P_{h_T}$ region, but, not enough to make up the order of magnitude difference~\cite{deFlorian:2013taa,Uebler:2017glm}.
Although the $P_{h_T}$-integrated cross section, which is dominated by the small-$P_{h_T}$ regime, has been well described within the collinear factorization, a successful description of the full $P_{h_T}$-distribution is critically important for understanding QCD dynamics and extracting the multidimensional partonic structure of the nucleon.  

The QCD factorization for high energy scattering cross sections with identified hadron(s), both TMD and collinear, is an approximation that neglects corrections suppressed by powers of the large momentum transfer of the hard collisions.  When $z_h\to 1$ in SIDIS, as shown in Eq.~\eqref{eq:fac-sidis}, the phase space for the fragmenting parton to shower (to radiate softer partons) is vanishing.  Consequently, the fragmentation function, $D_{j\to h}(z) \propto (1-z)^n \to 0$ at large $z$ with $n\ge 1$ for meson production (or a larger power for baryon production), strongly suppresses the probability for the produced active parton to become the observed hadron, and gives a powerlike suppression to the LP contribution to the SIDIS cross section in Eq.~\eqref{eq:fac-sidis}.  If next-to-leading power (NLP) contributions to the cross sections are factorizable, corresponding perturbatively calculable hard parts will be suppressed by inverse powers of the large momentum transfer, such as $1/Q^{\alpha}$ or $1/P_{h_T}^{\alpha}$ with $\alpha$ positive, in comparison with the hard parts of LP contributions.  However, the overall factorizable power suppressed contributions to the measured cross section could be sizable or even more important if the active partonic states, produced through the power suppressed short-distance partonic subprocesses, are much more likely to become the observed hadrons than the partonic states produced at the LP.  One example is the heavy quarkonium production at large transverse momentum at collider energies \cite{Kang:2011mg,Kang:2014tta,Kang:2014pya}, where producing a heavy quark pair at large $P_{h_T}$ is power suppressed comparing to the production of a single gluon at the same $P_{h_T}$, but, the probability for the produced heavy quark pair to become a bound quarkonium could be much greater than the probability for the produced gluon to become a heavy quarkonium in some kinematic regions.  

\begin{figure}[htp]
\centering
\includegraphics[width=0.18\textwidth]{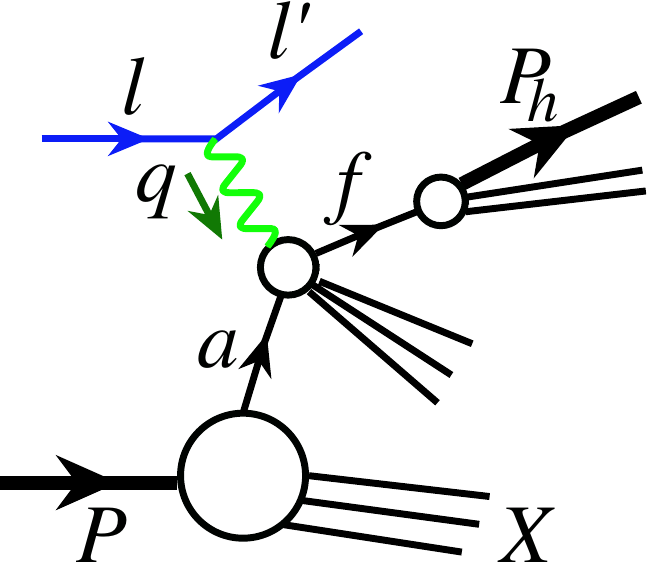}
\hskip 0.03\textwidth
\includegraphics[width=0.18\textwidth]{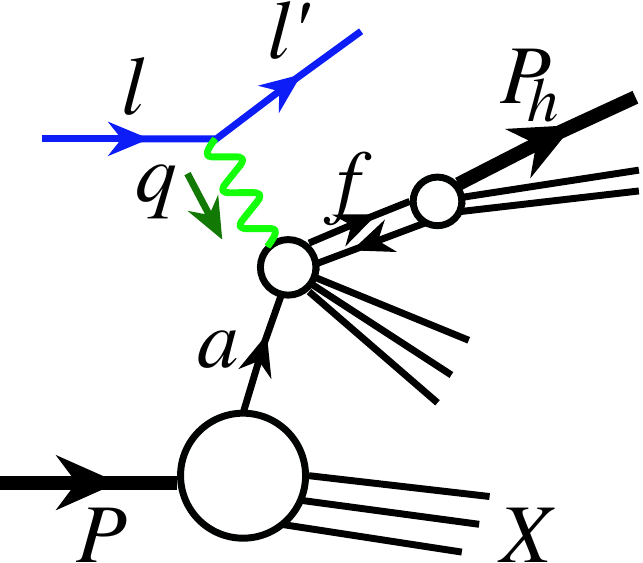}
\caption{Sample partonic channels for leading power (left) and next-to-leading power (right) contributions to lepton-hadron semi-inclusive DIS.}
\label{fig:lpnlp}
\end{figure}

For the production of a positively charged hadron in SIDIS, such as $\pi^+$, producing a quark-antiquark pair at high $P_{h_T}$, as sketched in Fig.~\ref{fig:lpnlp} (right), 
is certainly suppressed in comparison with the production of a single quark or antiquark at the same $P_{h_T}$, as shown in Fig.~\ref{fig:lpnlp}~(left).  However, a quark-antiquark pair with the right quantum number, {\it e.g.}, $u\bar{d}$ for $\pi^+$, could be much more likely to become the measured meson than a single quark or antiquark through the fragmentation process when the phase space for the radiation is vanishing and the multiplicity for the events is low. 
For example, when $z_h\to 1$, the factorized LP contribution to $\pi^+$ production in Eq.~\eqref{eq:fac-sidis} is suppressed by powers of $(1-z)$ from the FFs with $z\sim z_h$, and the factorizable NLP contribution, as we show in this paper, could have the leading transition from $u\bar{d}$ to $\pi^+$ to be proportional to $\delta(1-z)$ without the power suppression in $(1-z)$.  While the production of the $u\bar{d}$ pair is suppressed by inverse powers of $P_{h_T}$, as we will demonstrate below, the trade off between the $1/P_{h_T}^2$ suppressed hard parts at the NLP and the power suppressed FFs at the LP could make the formally power suppressed contributions to the SIDIS cross section very important for low multiplicity events.  

The importance of the power corrections to high energy pion production in SIDIS, as well as in $e^+e^-$ and hadron-hadron collisions, was recognized about 30 years ago in a series of publications by E.~Berger {\it et al.}~\cite{Berger:1979kz,Berger:1979xz,Berger:1980qg}.
In this paper, we investigate the NLP corrections to SIDIS production of charged mesons near the threshold, where $P_{h_T}\sim Q$ and $z_h\to 1$, in terms of QCD collinear factorization approach.  Instead of calculating all possible corrections at the NLP, we focus on the partonic subprocesses that could have the best chance to compete with the better studied LP contribution, and more specifically, we calculate the LO perturbative contribution in $\alpha_s$ to the production of a quark-antiquark pair that have the right flavor combination of the observed mesons.  In order to demonstrate the impact of the power corrections, quantitatively, we estimate the leading quark-antiquark pair FFs to a charged meson, which is proportional to $\delta(1-z)$, in terms of the better-known light-cone meson distribution amplitude square, by neglecting contributions suppressed by powers of $(1-z)$.  We find that the NLP corrections to SIDIS are extremely important for the production of charged mesons when the final-state multiplicity is low and the production kinematics is near the edge of phase space.  Our finding warrants a much more detailed study of power corrections to the SIDIS cross section near the edge of phase space where the hadron multiplicity is very low, which provides a unique opportunity to explore the mechanism of hadronization and color neutralization in QCD --the emergence of hadrons from produced quarks and gluons in high energy collisions.

The rest of this paper is organized as follows.  In Sec.~\ref{sec:nlp}, we introduce the factorization formalism for SIDIS to the accuracy of NLP, provide a leading order calculation of the short-distance hard parts at the NLP from the channels in which a quark-antiquark pair is produced with the same flavor combination of the valence content of the observed meson, and derive an approximate relation between the quark-antiquark FFs to a meson and the square of distribution amplitudes of the same meson.  In Sec.~\ref{sec:compare}, we show our numerical estimation of the size of the power corrections in comparison to the size of the LP contribution, quantitatively. Contributions from channels other than the direct production one are discussed in Sec.~\ref{sec:discussion}. Finally, conclusions and outlooks are given in Sec.~\ref{sec:summary}.

\section{Next-to-Leading Power Contribution to SIDIS \label{sec:nlp}}

With effectively one large momentum scale, $P_{h_T}\sim Q\gg \Lambda_{\rm QCD}$ observed, SIDIS cross section with a large $P_{h_T}$ hadron could be factorized in terms of the QCD collinear factorization approach.  With two identified hadrons, the initial-state hadron and the observed final-state hadron, only the leading and the first subleading power contributions to the SIDIS cross sections, in terms of the inverse power expansion of the observed large momentum transfer, are perturbatively factorizable, similar to the collinear factorization for inclusive Drell-Yan cross sections~\cite{Qiu:1990xy}.  In general, the subleading power contributions are much smaller comparing to the LP contributions because of the power suppression of the observed large momentum scale for the short-distance hard part, unless they can get enhancement from the hadronization~\cite{Kang:2011mg,Kang:2014tta,Kang:2014pya}, sufficient corrections to a steeply falling spectrum near the edge of available phase space~\cite{Qiu:1991pp,Qiu:1998ia}, or the multiple scattering in a large size and/or dense medium~\cite{Luo:1992fz,Qiu:2001hj,Guo:2000nz}.  Here, we consider possible enhancements from both the hadronization and the steep falling spectrum near the edge of phase space.

\subsection{The factorization formalism \label{sec:nlp-formalism}}

\begin{figure}[htp]
\centering
\includegraphics[width=0.22\textwidth]{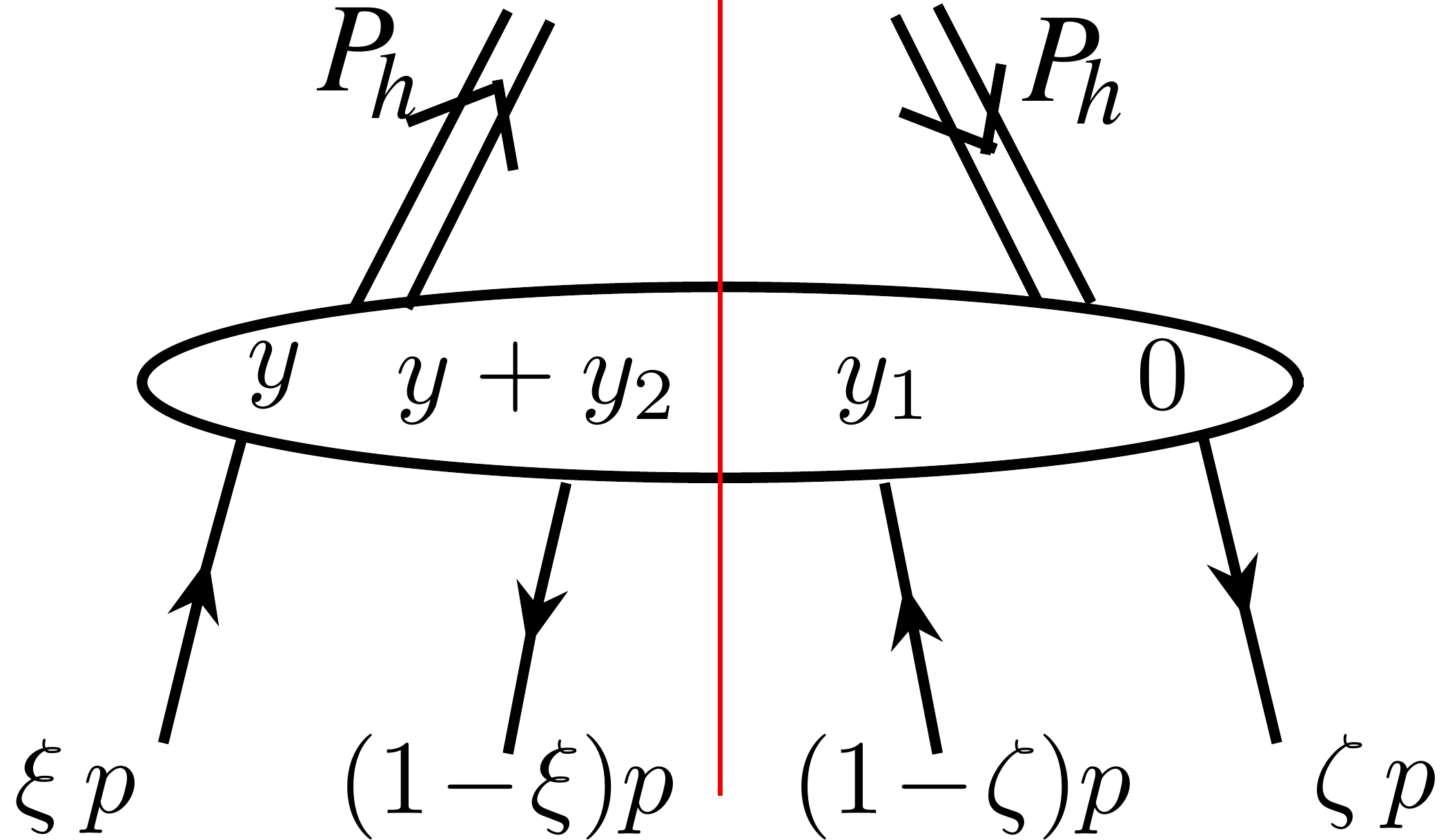}
\caption{Sketch of the fragmentation function for a quark-antiquark pair of momentum $p$ to fragment to a hadron of momentum $P_h$.}
\label{fig:2qFF}
\end{figure}

With the approximation of one-photon exchange, as shown in Fig.~\ref{fig:sidis}, the leptonic contributions to the SIDIS cross section is well understood and well defined.  In the rest of this paper, we present our calculations in terms of scattering of a virtual photon $\gamma^*$ of momentum $q$ with $Q^2=-q^2 > 0$ on a hadron $A$ of momentum $P$ to produce a charged hadron $h$ of momentum $P_h$: $\gamma^{*}(q)+A(P)\to h(P_h)+X$.  The corresponding formalisms can also cover the situation of photoproduction with a real photon $Q^2=0$.  When the photon is either deep virtual with $Q^2=-q^2 \sim P_{h_T}^2$ or real with $Q^2=0$, the scattering cross section can be expressed in terms of a collinearly factorized formalism \cite{Qiu:1990xy,Kang:2014tta},
\begin{widetext}
\begin{eqnarray}
\frac{d\sigma_{\gamma^{*}+A\to h+X}}{d^3{\bf P}_h/(2E_h)}
&\approx &
\sum_{a,f} \int_{x_B}^1 \frac{dx}{x} \int_{z_h}^1 \frac{dz}{z^2}\, \phi_{a/P}(x)\, D_{f\to h}(z) 
\frac{d\hat{\sigma}_{\gamma^{*}+a(l)\to f(p)+X}}{d^3{\bf p}/(2E_p)}
 \nonumber\\
&\ &
+\sum_{a,[ff'(\kappa)]} \int_{x_B}^1 \frac{dx}{x} \int_{z_h}^1 \frac{dz}{z^2} \int_0^1 d\xi d\zeta\, \phi_{a/P}(x)\, 
D_{[ff'(\kappa)]\to h}(z,\xi,\zeta)
\frac{d\hat{\sigma}_{\gamma^{*}+a(l)\to [ff'(\kappa)](p,\xi,\zeta)+X}}{d^3{\bf p}/(2E_p)},
\label{eq:factorize}
\end{eqnarray}
\end{widetext}
where $a$, $f$ (and $f'$) run over all parton flavors: $q$, $\bar{q}$, and $g$, $\kappa$ runs over all spin and color states of $[ff']$, $\phi_{a/P}(x)$ and $D_{f\to h}(z)$ are defined above as PDFs and FFs, respectively, with $l=xP$ and $P_h=zp$, and $D_{[ff'(\kappa)]\to h}(z,\xi,\zeta)$ are double-parton FFs, as sketched in Fig.~\ref{fig:2qFF}, with the momentum fraction of the pair carried by the observed hadron, $z=P_h/p$, and $\xi$ and $\zeta$ being relative momentum fractions of the two partons in the amplitude and its complex conjugate, respectively \cite{Kang:2011mg,Kang:2014tta,Kang:2014pya}.   
The factorization scale $\mu$ in Eq.~(\ref{eq:factorize}) is suppressed.  The $\hat{\sigma}_{\gamma^{*}+a(l)\to f(p)+X}$ and $\hat{\sigma}_{\gamma^{*}+a(l)\to [ff'(\kappa)](p,\xi,\zeta)+X}$ in Eq.~(\ref{eq:factorize}), respectively, are perturbatively calculable short-distance hard parts for producing a single parton of flavor $f$ and a pair of partons of flavor combination $[ff']$, while the hard part for producing a pair partons is suppressed by $1/P_{h_T}^2$ in comparison with the hard part for producing a single parton at the same momentum $p$.  In Eq.~(\ref{eq:factorize}), we neglected other NLP terms proportional to twist-4 parton distributions and the same single parton FFs, $\phi^{(4)}_{[ab]/P}(x,x',x'')\otimes D_{f\to h}(z)$, since corresponding partonic hard part is power suppressed comparing the LP hard part, while no potential enhancement from the hadronization since the same FFs are used.   There could be additional contributions to the factorized formalism in Eq.(\ref{eq:factorize}), and they are typically suppressed by even higher power in $1/P_{h_T}^2$ and/or $1/Q^2$.  These contributions are not expected to be factorizable.  The partonic short-distance hard parts in Eq.~(\ref{eq:factorize}) at the leading order in power of strong coupling constant $\alpha_s$ are given by
\begin{eqnarray}
\frac{E_p d\hat{\sigma}_{\gamma^{*}+a(l)\to f(p)+X}}{d^3{\bf p}}
&=&
\frac{|\overline{\cal M}_{\gamma^{*}+a(l)\to f(p)+X}|^2}{2(\hat{s}+Q^2)}
\nonumber \\
&\ & {\hskip -0.3in} \times
\frac{1}{2(2\pi)^2} \delta(\hat{s}+\hat{t}+\hat{u}+Q^2),
\label{eq:hard-lp}
\\
\frac{E_p d\hat{\sigma}_{\gamma^{*}+a(l)\to [ff'](p)+X}}{d^3{\bf p}}
&=&
\frac{|\overline{\cal M}_{\gamma^{*}+a(l)\to [ff'(\kappa)](p)+X}|^2}{2(\hat{s}+Q^2)} 
\nonumber \\
&\ & {\hskip -0.3in} \times
\frac{1}{2(2\pi)^2}\delta(\hat{s}+\hat{t}+\hat{u}+Q^2),
\label{eq:hard-nlp}
\end{eqnarray}
where parton level Mandelstam variables are defined as
\begin{align}
\hat{s}=(q+l)^2,
\quad
\hat{t}=(q-p)^2,
\quad
\hat{u}=(l-p)^2,
\end{align}
with the constraint 
\begin{align}
\hat{s}+\hat{t}+\hat{u}=q^2=-Q^2
\end{align}
imposed by the phase space $\delta$-function of the massless two-particle final-state and the momentum conservation of the $2\to 2$ partonic subprocess.
In Eqs.~(\ref{eq:hard-lp}) and (\ref{eq:hard-nlp}), the denominator $2(\hat{s}+Q^2) = 2E_{\gamma^*} 2E_a |v_{\gamma^*}-v_a|$ is the flux factor of the partonic scattering process, and $|\overline{\cal M}_{\gamma^{*}+a(l)\to f(p)+X}|^2$ and $|\overline{\cal M}_{\gamma^{*}+a(l)\to [ff'(\kappa)](p)+X}|^2$, respectively, are squared LP and NLP partonic scattering amplitudes with initial-state spin and color averaged and final-state spin and color summed. 

With one large momentum transfer, $Q^2$ or $P_{h_T}^2$, in the scattering involving two identified hadrons, the collinear factorization of the leading power term and the {\it first} subleading power term, which correspond to the first and second terms on the right-hand side of Eq.~\eqref{eq:factorize}, respectively, is valid to the same level of arguments (or proof) in QCD perturbation theory~\cite{Qiu:1990xy,Kang:2014tta}.  However, our knowledge of the second term in Eq.~\eqref{eq:factorize}, or power corrections in general, is much less than the first term, while the second term could help reveal additional and richer QCD dynamics, such as color entanglement and parton-parton correlations.  The new and precise data from COMPASS and future experiments at Jefferson Lab, and the apparent discrepancy from the LP predictions could provide new opportunities to study and explore the rich QCD dynamics at the NLP, while its exploratory study was initiated over 30 years ago.

\subsection{Power suppressed partonic hard parts \label{sec:nlp-hard}}

The theoretical prediction (or an estimation) for the size of power corrections in Eq.~\eqref{eq:factorize} relies on our ability to calculate the short-distance hard parts to produce a pair of partons and our knowledge of the nonperturbative double-parton FFs. For getting the most enhancement from the pair's hadronization to compensate the power suppression of the partonic hard parts to produce the pair at NLP, we calculate the partonic hard parts from subprocesses that can produce a quark-antiquark pair with the same valence quark flavor combination as the produced charged meson.  

\begin{figure}[htp]
\centering
\includegraphics[width=0.2\textwidth]{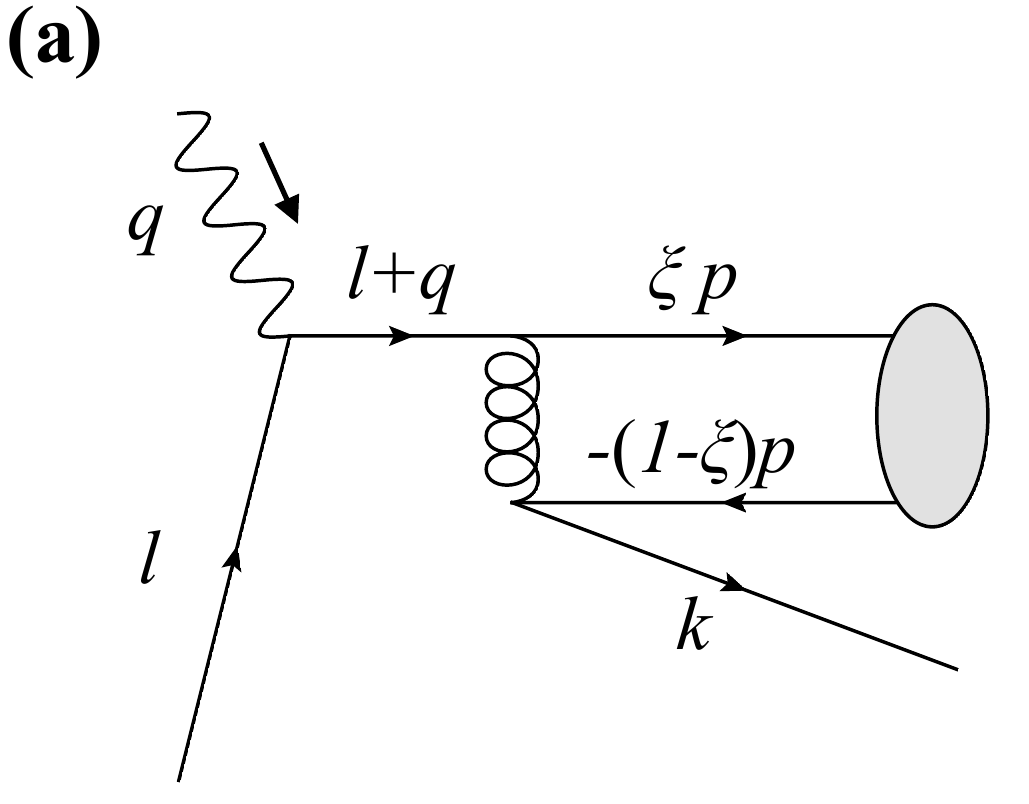}\ \
\includegraphics[width=0.2\textwidth]{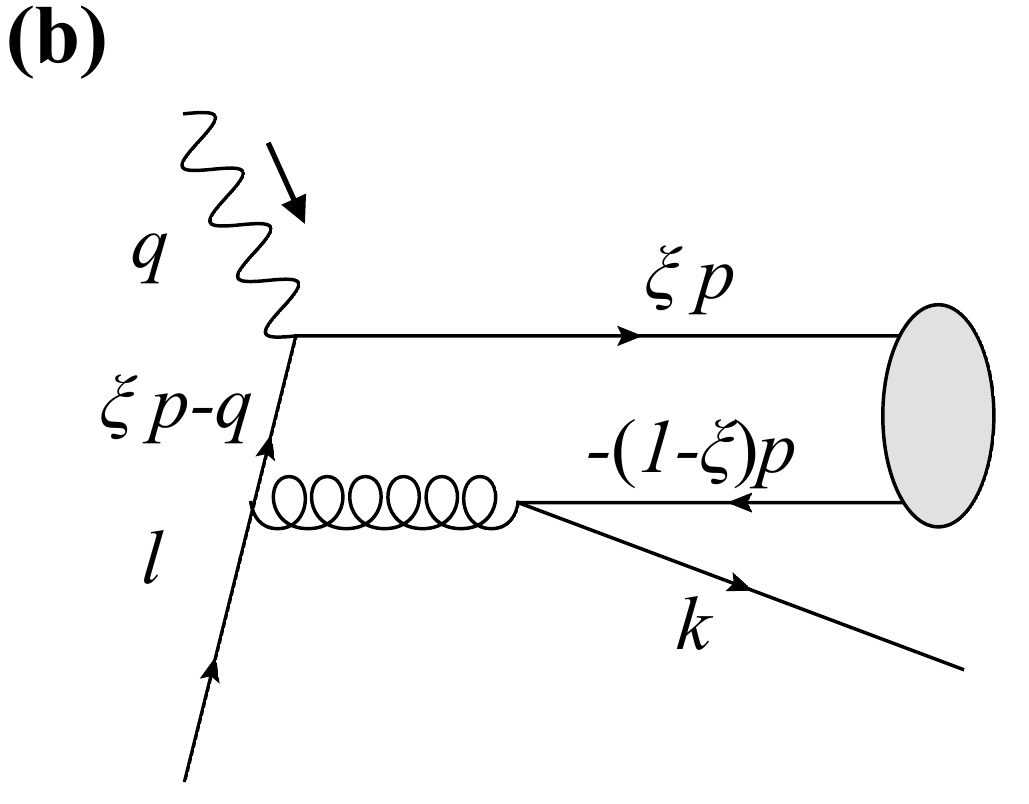}
\includegraphics[width=0.2\textwidth]{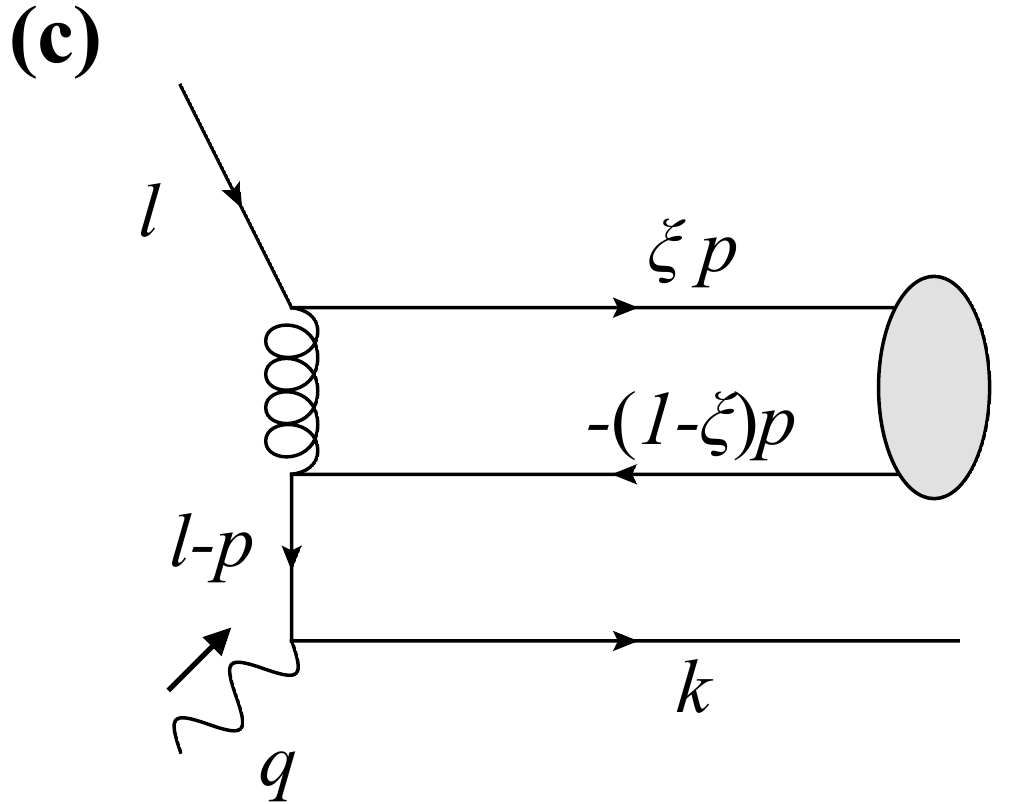}\ \
\includegraphics[width=0.2\textwidth]{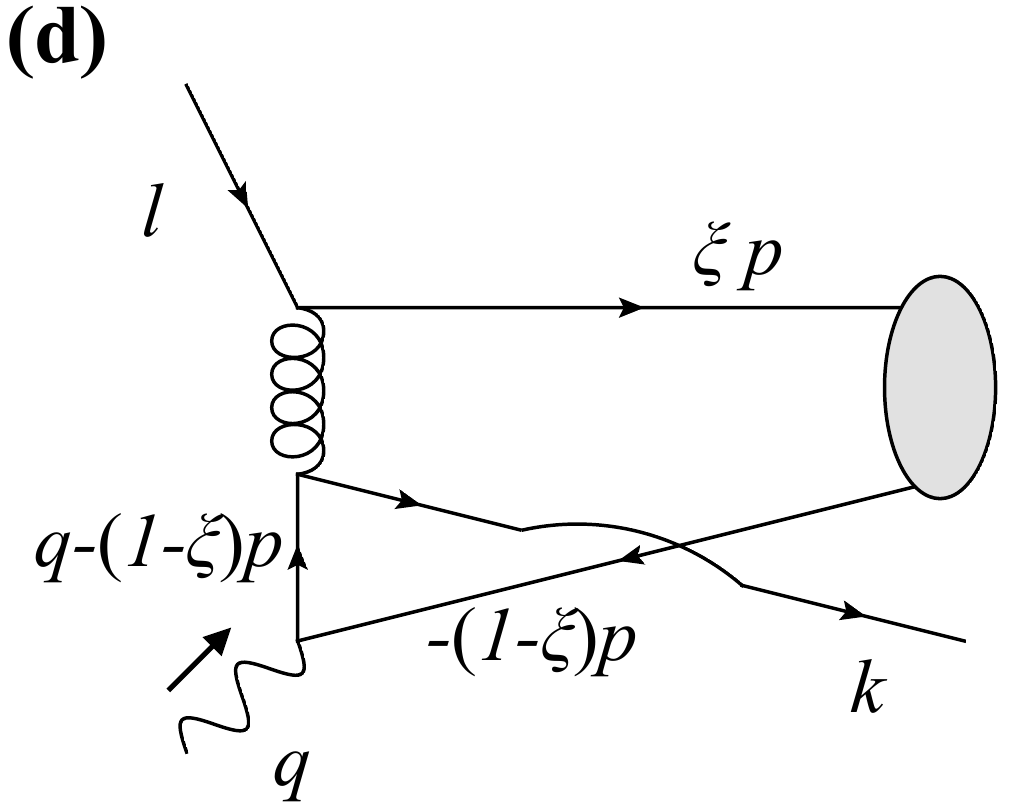}
\caption{Leading order Feynman diagrams for the scattering amplitude of partonic subprocess:\ $\gamma^{*}+q\to [q\bar{q}'(\kappa)]+q'$.}
\label{feyn}
\end{figure}

At the lowest order in power of $\alpha_s$, we need to consider the subprocess:\  $\gamma^{*}+q\to [q\bar{q}'(\kappa)]+q'$, with corresponding Feynman diagrams given in Fig.~\ref{feyn} where the quark and antiquark flavors $q$ and $\bar{q}'$ should match the valence flavors of the measured meson.  The color state of the produced quark-antiquark pair can be either a singlet ``$[1]$'' or an octet ``$[8]$'' state, and corresponding color projection operators are respectively proportional to $\delta_{ab}$ and $t^A_{ab}$, where $a,b=1,2, ..., N_c$ are color indices for the quark and the antiquark, and $t^A_{ab}$ is the generator of the $SU(N_c)$ color group in the fundamental representation with $A=1,2, ..., N_c^2-1$ and $N_c=3$ in QCD. In this paper, we use the same color projection operators for the hard part as done in Refs.~\cite{Kang:2014tta,Kang:2014pya},
\begin{align}
\tilde{\cal C}^{[1]}_{ba,dc}&=\delta_{ba}\delta_{dc},\\
\tilde{\cal C}^{[8]}_{ba,dc}&=\sum_A \sqrt{2}t^A_{ba}\sqrt{2}t^A_{dc},
\end{align}
where $d,c$ are color indices for the quark-antiquark pair in the complex-conjugate of the scattering amplitudes.
The corresponding color projection operators for the quark-antiquark FFs in Fig.~\ref{fig:2qFF}
are \cite{Kang:2014tta,Kang:2014pya}
\begin{align}
{\cal C}^{[1]}_{ab,cd}&=\frac{1}{N_c^2}\delta_{ab}\delta_{cd},\\
{\cal C}^{[8]}_{ab,cd}&=\frac{1}{N_c^2-1}\sum_A \sqrt{2}t^A_{ab}\sqrt{2}t^A_{cd}.
\end{align}
They satisfy the normalization condition,
\begin{align}
\sum_{abcd} \tilde{\cal C}^{I}_{ba,dc} {\cal C}^{J}_{ab,cd}=\delta^{IJ},
\end{align}
where $I,J=[1],[8]$.

The spin projection operators for the four spin states of the quark-antiquark pair can be given by $(\gamma\cdot p)_{ij}$,\break 
$(\gamma\cdot p\,\gamma_5)_{ij}$, and $(\gamma\cdot p\,\gamma_\perp^\alpha)_{ij}$ with $\alpha=1,2$, or their linear combinations. They could be referred to as the vector $(v)$, axial-vector $(a)$, and tensor $(t)$ projections. Following Refs.~\cite{Kang:2014tta,Kang:2014pya}, we choose the spin projection operators for the hard part as
\begin{align}
\tilde{\cal P}^{(v)}(p)_{ji,lk}&=(\gamma\cdot p)_{ji} (\gamma\cdot p)_{lk},\\
\tilde{\cal P}^{(a)}(p)_{ji,lk}&=(\gamma\cdot p\, \gamma_5)_{ji} (\gamma\cdot p\, \gamma_5)_{lk},\\
\tilde{\cal P}^{(t)}(p)_{ji,lk}&=\sum_{\alpha=1,2}(\gamma\cdot p\, \gamma_\perp^\alpha)_{ji}(\gamma\cdot p\, \gamma_\perp^\alpha)_{lk},
\end{align}
and corresponding spin projection operators for the quark-antiquark FFs to be
\begin{align}
{\cal P}^{(v)}(p)_{ij,kl}&=\frac{1}{4p\cdot n}(\gamma\cdot n)_{ij} \frac{1}{4p\cdot n}(\gamma\cdot n)_{kl},\\
{\cal P}^{(a)}(p)_{ij,kl}&=\frac{1}{4p\cdot n}(\gamma\cdot n\, \gamma_5)_{ij} \frac{1}{4p\cdot n}(\gamma\cdot n\, \gamma_5)_{kl},\\
{\cal P}^{(t)}(p)_{ij,kl}&=\frac{1}{2}\sum_{\alpha=1,2}\frac{1}{4p\cdot n}(\gamma\cdot n\, \gamma_\perp^\alpha)_{ij}\frac{1}{4p\cdot n}(\gamma\cdot n\, \gamma_\perp^\alpha)_{kl},
\end{align}
where $n$ is a null vector with $n^2=0$, defined to be conjugated to the momentum of the quark-antiquark pair $p$, such that $p\cdot n$ is the only nonvanishing component of $p^\mu$ if $p^2 = 0$. As required, the spin projection operators satisfy the normalization condition,
\begin{align}
\sum_{ijkl} \tilde{\cal P}^{(s)}_{ji,lk} {\cal P}^{(s')}_{ij,kl}=\delta^{ss'},
\end{align}
where $s,s'=v,a,t$.

The spin state of a virtual photon of momentum $q$ can be either transversely polarized with the polarization vectors $\epsilon_\pm^\mu$ or longitudinally polarized with the polarization vector $\epsilon_L^\mu$. The transverse spin polarization tensor is defined as
\begin{align}
d^{\mu\nu}(q)&=\sum_{\lambda=\pm}\epsilon_\lambda^{*\mu}\epsilon_\lambda^\nu
=-g^{\mu\nu} + v^\mu \bar{v}^\nu + \bar{v}^\mu v^\nu,
\label{eq:proj_T}
\end{align}
and the longitudinal spin polarization tensor is given by
\begin{align}
K^{\mu\nu}(q)=\epsilon_L^{*\mu}\epsilon_L^\nu
=&\frac{1}{-q^2}\left[(q\cdot \bar{v})^2 v^\mu v^\nu 
+ (q\cdot v)^2 \bar{v}^\mu \bar{v}^\nu
\right]
\nonumber \\
& + \frac{1}{2}(v^\mu \bar{v}^\nu + \bar{v}^\mu v^\nu),
\label{eq:proj_L}
\end{align}
where $v$ and $\bar{v}$ are two null vectors introduced to pick the ``$+$'' and ``$-$'' light-cone components of the photon momentum $q$ with $v^2=
\bar{v}^2=0$, $v\cdot \bar{v}=1$, and $q^\mu=(q\cdot v)\bar{v}^\mu+(q\cdot\bar{v})v^\mu$.

The color factors for all the squares of diagrams in Fig.~\ref{feyn} are the same for each color projection:
\begin{align}
C^{[1]}&=\frac{1}{N_c}\sum_{AB}{\rm Tr}[t^A t^A t^B t^B]=\frac{(N_c^2-1)^2}{4N_c^2},\\
C^{[8]}&=\frac{1}{N_c}\sum_{ABC}2{\rm Tr}[t^A t^C t^A t^B t^C t^B]=\frac{N_c^2-1}{4N_c^3},
\end{align}
where the factor $1/N_c$ is from the average of initial state quark colors. Thus we can factor it out from amplitude squares of the diagrams in Fig.~\ref{feyn}. 

\begin{widetext}
Now, we calculate the invariant scattering amplitude squares from the diagrams in Fig.~\ref{feyn} for the production of a quark-antiquark pair in the axial-vector spin state.  For example, the invariant amplitude square of the diagram (a) in Fig.~\ref{feyn} with transversely polarized photon is given by
\begin{equation}
\begin{split}
\big|\overline{\cal M}^{a^\dagger a}_T\big|^2 
&=
\frac{1}{2} C e^2 e_q^2 g_s^4 {\rm Tr}\left[\slashed{l}\gamma^\mu(\slashed{l} + \slashed{q})\gamma^\alpha\slashed{p}\gamma_5
\gamma^\beta(\slashed{l}+\slashed{q}-\slashed{p})
\gamma^\sigma\slashed{p}\gamma_5\gamma^\rho(\slashed{l}+\slashed{q})\gamma^\nu\right]\\
&\quad\times\frac{1}{(l+q)^2}\frac{1}{(l+q)^2}
\frac{-g_{\alpha\beta}}{(l+q-\zeta p)^2}\frac{-g_{\rho\sigma}}{(l+q-\xi p)^2}\frac{d_{\mu\nu}(q)}{2},
\end{split}
\label{eq:maa}
\end{equation}
where the factor $1/2$ is from the spin average of the initial state quark, $C=C^{[1]}$ (or $C^{[8]}$) is the color factor, $d^{\mu\nu}(q)/2$ projects out one transversely polarized state of the colliding photon of momentum $q$, and Feynman gauge was used for the gluon propagators. For calculating contributions from the scattering of a longitudinally polarized virtual photon, one only needs to replace $d_{\mu\nu}(q)/2$ above by $K_{\mu\nu}(q)$ defined in Eq.~(\ref{eq:proj_L}). We obtain the invariant amplitude squares of all diagrams in Fig.~\ref{feyn} with a transversely polarized photon and the produced quark-antiquark pair in an axial-vector spin state,  
\begin{align}
\big|\overline{\cal M}_T^{a^\dagger a + a^\dagger b + b^\dagger a + b^\dagger b}\big|^2
&=
4\, C K e_q^2
\left[
-\frac{(\hat{t}+2\hat{u})A_{\bar{\xi}}A_{\bar{\zeta}}+\hat{u}^2(\hat{t}-\bar{\xi}A_{\bar{\zeta}}-\bar{\zeta}A_{\bar{\xi}})}{\bar{\xi}\bar{\zeta}\hat{s}^2 A_{\bar{\xi}}A_{\bar{\zeta}}}
+\frac{2\hat{u}^3(\hat{s}+\hat{t}+\hat{u})}{\hat{s}(\hat{t}+\hat{u})^2A_{\bar{\xi}}A_{\bar{\zeta}}}
\right],
\label{eq:absquare:t}\\
\big|\overline{\cal M}_T^{a^\dagger c + a^\dagger d + b^\dagger c + b^\dagger d}\big|^2
&=
4\, C K e_q e_{q'}
\left[\frac{(\hat{t}+2\hat{u})A_{\xi}A_{\bar{\zeta}}+B_\xi A_{\bar{\zeta}}-\bar{\zeta}\hat{u}^2A_\xi}{\xi\bar{\zeta}\hat{s}\hat{u}A_\xi A_{\bar{\zeta}}}
-\frac{2\hat{u}(\hat{s}+\hat{t}+\hat{u})(\hat{t}+\xi\hat{u})}{\xi(\hat{t}+\hat{u})^2 A_\xi A_{\bar{\zeta}}}
\right],\\
\big|\overline{\cal M}_T^{c^\dagger a + c^\dagger b + d^\dagger a + d^\dagger b}\big|^2
&=
4\, C K e_q e_{q'}
\left[\frac{(\hat{t}+2\hat{u})A_{\bar{\xi}}A_{\zeta}+B_\zeta A_{\bar{\xi}}-\bar{\xi}\hat{u}^2A_\zeta}{\bar{\xi}\zeta\hat{s}\hat{u}A_{\bar{\xi}} A_{\zeta}}
-\frac{2\hat{u}(\hat{s}+\hat{t}+\hat{u})(\hat{t}+\zeta\hat{u})}{\zeta(\hat{t}+\hat{u})^2 A_{\bar{\xi}} A_{\zeta}}
\right],\\
\big|\overline{\cal M}_T^{c^\dagger c + c^\dagger d + d^\dagger c + d^\dagger d}\big|^2
&=
4\, C K e_{q'}^2
\left[-\frac{(\hat{t}+2\hat{u})A_\xi A_\zeta + B_\xi A_\zeta + B_\zeta A_\xi + \hat{s}^2\hat{t}}{\xi\zeta\hat{u}^2A_\xi A_\zeta}
+\frac{2\hat{s}(\hat{s}+\hat{t}+\hat{u})(\hat{t}+\xi\hat{u})(\hat{t}+\zeta\hat{u})}{\xi\zeta\hat{u}(\hat{t}+\hat{u})^2A_\xi A_\zeta}
\right],
\end{align}
where the parameters are defined as 
\begin{align}
K &= e^2 g_s^4 = (4\pi)^3 \alpha_{\rm em} \alpha_s^2,
\label{eq:para-K}\\
A_\eta &= \hat{t} + \eta (\hat{s} + \hat{u}),\\
B_\eta &= \hat{s} \hat{t} - \hat{u}(\hat{t} + \eta \hat{u})
\label{eq:para-B},
\end{align}
with $\eta=\xi,\zeta,\bar{\xi},\bar{\zeta}$ and $\bar{\xi}=1-\xi$, $\bar{\zeta}=1-\zeta$, respectively. 
Similarly, for a longitudinally polarized photon, we have 
\begin{align}
|\overline{\cal M}_L^{a^\dagger a + a^\dagger b + b^\dagger a + b^\dagger b}|^2
& =
\frac{16\, C K e_q^2 \hat{u}^3 (\hat{s}+\hat{t}+\hat{u})}{\hat{s}(\hat{t}+\hat{u})^2 A_{\bar{\xi}} A_{\bar{\zeta}}},\\
|\overline{\cal M}_L^{a^\dagger c + a^\dagger d + b^\dagger c + b^\dagger d}|^2
& =
\frac{-16\, C K e_q e_{q'} \hat{u}(\hat{t}+\xi\hat{u})(\hat{s}+\hat{t}+\hat{u})}{\xi(\hat{t}+\hat{u})^2 A_{\xi} A_{\bar{\zeta}}},\\
|\overline{\cal M}_L^{c^\dagger a + c^\dagger b + d^\dagger a + d^\dagger b}|^2
& =
\frac{-16\, C K e_q e_{q'} \hat{u}(\hat{t}+\zeta\hat{u})(\hat{s}+\hat{t}+\hat{u})}{\zeta(\hat{t}+\hat{u})^2 A_{\bar{\xi}} A_{\zeta}},\\
|\overline{\cal M}_L^{c^\dagger c + c^\dagger d + d^\dagger c + d^\dagger d}|^2
& =
\frac{16\, C K e_{q'}^2 \hat{s}(\hat{t}+\xi\hat{u})(\hat{t}+\zeta\hat{u})(\hat{s}+\hat{t}+\hat{u})}{\xi\zeta\hat{u}(\hat{t}+\hat{u})^2 A_{\xi} A_{\zeta}},
\label{eq:cdsquare:l}
\end{align}
where all parameters are the same as those defined in Eqs.~(\ref{eq:para-K}) - (\ref{eq:para-B}).
\end{widetext}

By replacing the axial-vector spin projection for the produced quark-antiquark pair by a vector spin projection, we find that the invariant amplitude squares of all diagrams in Fig.~\ref{feyn} have the same results as those from Eqs.~(\ref{eq:absquare:t}) to (\ref{eq:cdsquare:l}).  This is easy to understand at this order of calculation since we neglect the quark mass and the two $\gamma_5$ in the spin trace, as those in Eq.~(\ref{eq:maa}), can be combined into $\gamma_5^2=1$.

\subsection{The quark-antiquark fragmentation function \label{sec:nlp-fragmentation}}

In order to quantitatively estimate the size of the power corrections to the semi-inclusive production of large $P_{h_T}$ hadrons, in terms of the factorization approach in Eq.~(\ref{eq:factorize}), we need the knowledge of quark-antiquark FFs in addition to the calculation of perturbative partonic hard parts.  For getting the most enhancement from the fragmentation of the quark-antiquark pair, we focus on the production channels in which the flavor of the produced quark-antiquark pair $[ff']$ matches the flavor combination of the valence components, $[q\bar{q}']$, of the produced meson $h$.  As introduced in Ref.~\cite{Kang:2014tta}, the quark-antiquark FFs $D_{[q\bar{q}'(\kappa)]\to h}(z,\xi,\zeta)$ are defined as
\begin{align}
&D_{[q\bar{q}'(\kappa)]\to h}(z,\xi,\zeta)
=\sum_X 
\int \frac{P_h^+ dy^-}{2\pi}\!\!
\int \frac{P_h^+ dy_1^-}{2\pi}\!\!
\int \frac{P_h^+ dy_2^-}{2\pi}
\nonumber\\
&\quad\times
e^{i(1-\zeta)\frac{P_h^+}{z}y_1^-}
e^{-i\frac{P_h^+}{z}y^-}
e^{-i(1-\xi)\frac{P_h^+}{z}y_2^-}
\label{eq:qq2piFF}\\
&\quad\times 
{\cal C}\, {\cal P}\,
\langle 0|\bar{q}'(y_1^-) [\Phi_n(y_1^-)]^\dagger [\Phi_n(0)] q(0) |h(P_h)X\rangle
\nonumber\\
&\quad \times
\langle h(P_h)X|\bar{q}(y^-) [\Phi_n(y^-)]^\dagger [\Phi_n(y^-\!\! +y_2^-)] q'(y^-\!\! +y_2^-) |0\rangle,
\nonumber
\end{align}
where ${\cal C}$ and ${\cal P}$ are respectively the color and spin projection operators defined in Sec.~\ref{sec:nlp-hard} with color and spin indices suppressed.  The $\Phi_n(y^-)$ in Eq.~(\ref{eq:qq2piFF}) is the gauge link in the fundamental representation of QCD color group, given by
\begin{align}
\Phi_n(y^-)&={ P} \exp\left[ -ig_s\int_{y^-}^{\infty} d\lambda\, n\cdot G^A(\lambda n)\, t^A\right],
\end{align}
where ${P}$ and $G^A$ represent the path ordering and the gluon field, respectively, and $t^A$ is the generator of SU(3) color with color index $A$, as introduced in Sec.~\ref{sec:nlp-hard}.

Like the single-parton FFs, the quark-antiquark FFs are nonperturbative and cannot be calculated within the QCD perturbation theory, while their factorization scale dependence could be calculated if the variation is within the perturbative regime~\cite{Kang:2014tta}.  In order to estimate the size of the power corrections, quantitatively, we make the following approximation.  We assume that at an input scale $\mu_0$, the quark-antiquark pair FFs are dominated by the final-state in which there is no additional hadron produced other than the observed meson, which mimics the physical condition when the observed meson is produced near the edge of phase space with very low multiplicity.  That is, the hadronic final-state in Eq.~\eqref{eq:qq2piFF} is approximated as $|h(P_h)X\rangle \approx |h(P_h)\rangle$. 
Under this approximation, as shown below, we can relate the quark-antiquark FFs to the square of meson distribution amplitude, to help us to estimate the size (possibly a lower limit) of the power corrections quantitatively by using the better known knowledge on the meson distribution amplitudes.  

For pseudoscalar mesons, {\it e.g.}, pions and kaons, the distribution amplitude $\phi_h(x,\mu)$ has the following matrix element definition,
\begin{align}
& {\hskip -0.1in}
\langle 0|\bar{q}_{a,i}(y^-\!\! +y_1^-)(\gamma\! \cdot\! n \gamma_5)_{ij} U_{ab}
(y^-\!\! +y_1^-,y^-) q_{b,j}(y^-)|h(P_h)\rangle
\nonumber\\
&\quad =
iP_h^+ f_h \int_0^1 dx\,
e^{-ixP_h^+ y^- -i(1-x)P_h^+(y^-+y_1^-)}
\phi_h(x,\mu)
\nonumber \\
&\quad =
iP_h^+ f_h\,  e^{-iP_h^+y^-}\!\! \int_0^1 dx\, 
e^{-i(1-x)P_h^+y_1^-}\phi_h(x,\mu),
\label{eq:da}
\end{align}
where the contraction of color and spin indices are explicitly shown, $U_{ab}(y_2^-,y_1^-)=[\Phi_n(y_2^-)]^\dagger_{ac}[\Phi_n(y_1^-)]_{cb}$, $f_h$ is the decay constant of the meson $h$, and $\phi_h(x,\mu)$ is the meson's distribution amplitude. 

\begin{widetext}
From the definition in Eq.~(\ref{eq:qq2piFF}), we can rewrite the double-parton FFs of a quark-antiquark pair with color singlet and axial-vector projections at the input scale $\mu_0$ as follows,
\begin{align}
D_{[q\bar{q}'(1a)]}(z,\xi,\zeta,\mu_0) &\approx 
\int \frac{P_h^+dy^-}{2\pi}
\int \frac{P_h^+dy_1^-}{2\pi}
\int \frac{P_h^+dy_2^-}{2\pi}
e^{i(1-\zeta)\frac{P_h^+}{z}y_1^-}
e^{-i\frac{P_h^+}{z}y^-}
e^{-i(1-\xi)\frac{P_h^+}{z}y_2^-}
\nonumber\\
&\quad \times
\frac{1}{4N_c P_h^+}\langle 0| \bar{q}'_{c',k}(y_1^-)(\gamma\cdot n\gamma_5)_{kl}U_{c'd'}(y_1^-,0) q_{d',l}(0) |h(P_h)\rangle
\nonumber\\
&\quad \times
\frac{1}{4N_c P_h^+}\langle h(P_h)| \bar{q}_{a',i}(y^-)(\gamma\cdot n\gamma_5)_{ij}U_{a'b'}(y^-,y^-+y_2^-) q'_{b',j}(y^-+y_2^-) |0\rangle
\label{eq:operator-def}\\
&=\frac{1}{16N_c^2}
\int \frac{P_h^+dy^-}{2\pi}
\int \frac{P_h^+dy_1^-}{2\pi}
\int \frac{P_h^+dy_2^-}{2\pi}
e^{i(1-\zeta)\frac{P_h^+}{z}y_1^-}
e^{-i\frac{P_h^+}{z}y^-}
e^{-i(1-\xi)\frac{P_h^+}{z}y_2^-}
\nonumber\\
&\quad \times
f_h^2\, e^{iP_h^+y^-}\int_0^1 d\zeta' e^{-i(1-\zeta')P_h^+ y_1^-}\phi_h(\zeta',\mu_0)
\int_0^1 d\xi' e^{i(1-\xi')P_h^+y_2^-}\phi_h(\xi',\mu_0)
\nonumber\\
&=\frac{f_h^2}{16N_c^2}\, z\, \delta(1-z)
\phi_h(\zeta,\mu_0)\phi_h(\xi,\mu_0).
\label{eq:qqbarFFs}
\end{align}
\end{widetext}
In deriving the above result, the relation in Eq.~(\ref{eq:da}) was used.  Under this extreme approximation, $|h(P_h)X\rangle \approx |h(P_h)\rangle$,
the nonperturbative quark-antiquark pair FFs can be expressed in terms of a product of two nonperturbative meson distribution amplitudes $\phi_h(\zeta,\mu_0)$ and $\phi_h(\xi,\mu_0)$, which have been better studied than the quark-antiquark pair FFs.   We will come back to discuss the corrections to our extreme approximation in Sec.~\ref{sec:discussion}.

\section{Comparison between LP and NLP contributions \label{sec:compare}}

In order to understand the relevance and potential impact of the NLP corrections, we evaluate and compare the size of the LP and NLP contributions to the SIDIS production of a charged meson at large transverse momentum, numerically, in this section. 

The differential multiplicities for charged hadrons in lepton DIS off a deuteron target were recently measured by the COMPASS Collaboration~\cite{Aghasyan:2017ctw}. The differential multiplicity is defined as the ratio between the SIDIS and the inclusive DIS differential cross sections:
\begin{align}
\frac{d^2M_h}{dz_h dP_{h_T}^2}&=
\left(\frac{d^4\sigma^{\rm SIDIS}_h}{dx_B dQ^2 dz_h dP_{h_T}^2}\right)
\bigg/
\left(\frac{d^2\sigma^{\rm DIS}}{dx_B dQ^2}\right),
\label{eq:multiplicity}
\end{align}
where $x_B$ and $z_h$ are defined in Eq.~(\ref{eq:xbzh}), and $P_{h_T}$ is defined in the photon-target frame. Since the produced hadrons are dominated by pions, we only calculate SIDIS cross sections for charged pions, $\pi^\pm$ in this section, instead of the sum of all long-lived charged hadrons, $h^\pm$, as included in the COMPASS data. 

\begin{figure}[htp]
\centering
\includegraphics[width=0.23\textwidth]{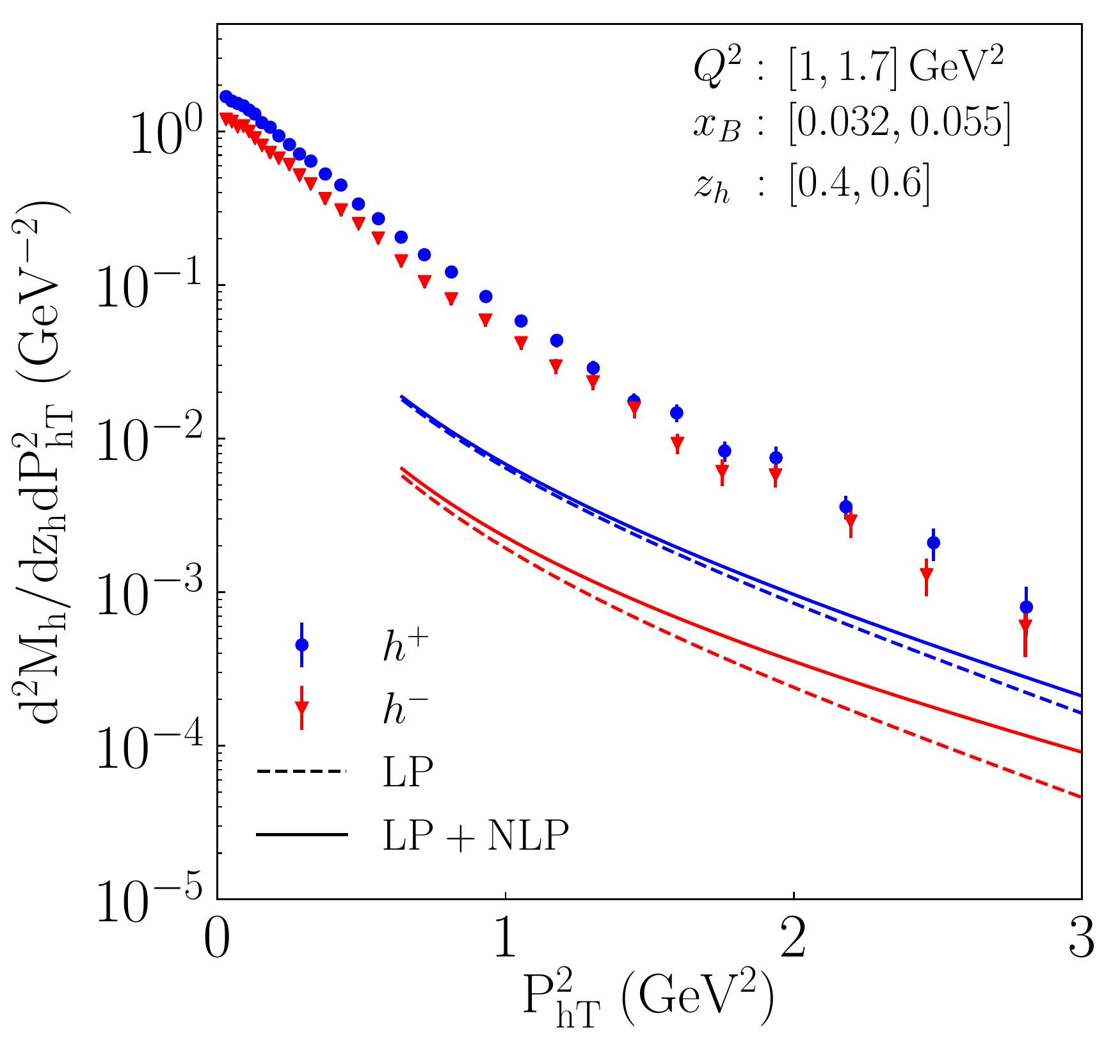}
\includegraphics[width=0.23\textwidth]{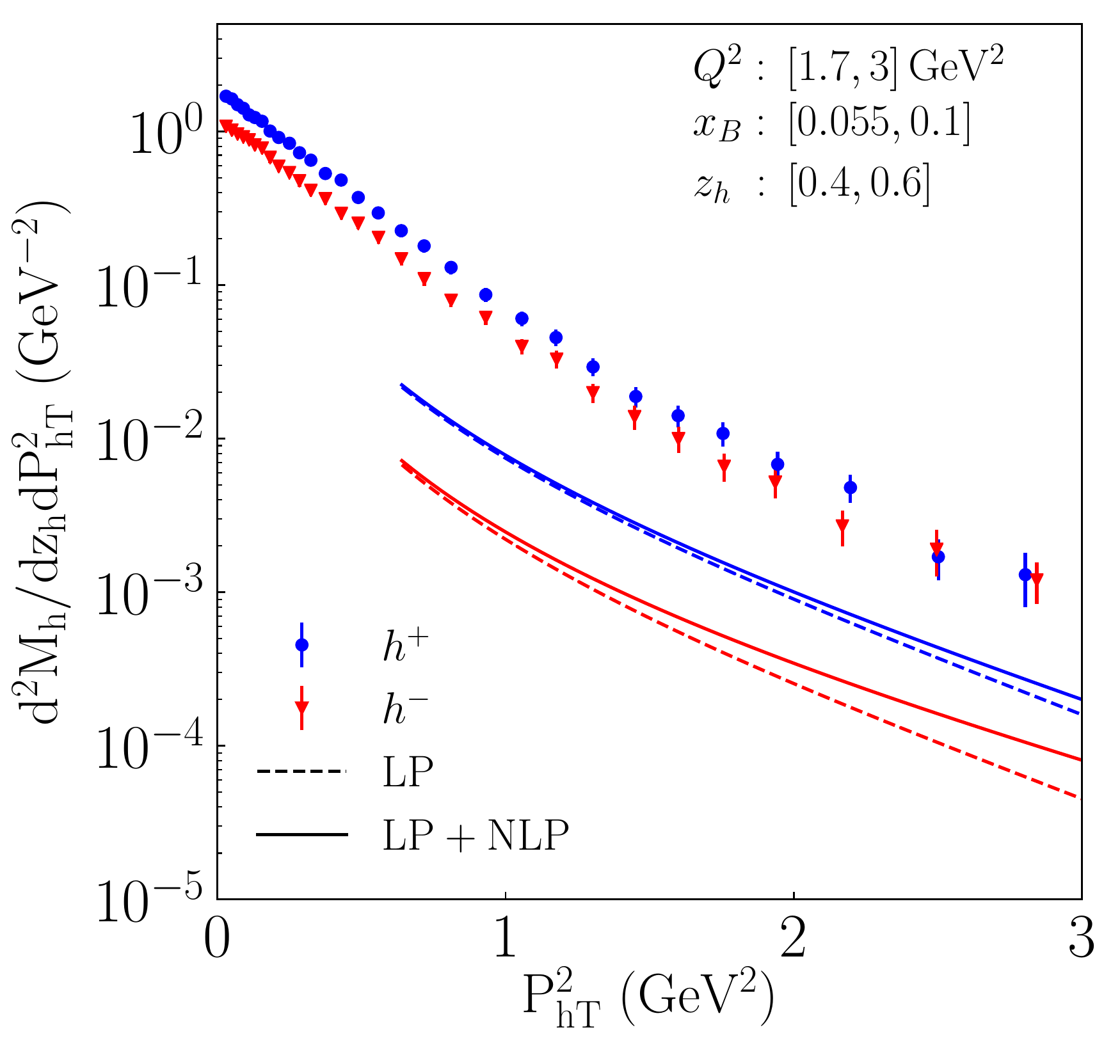}\\
\includegraphics[width=0.23\textwidth]{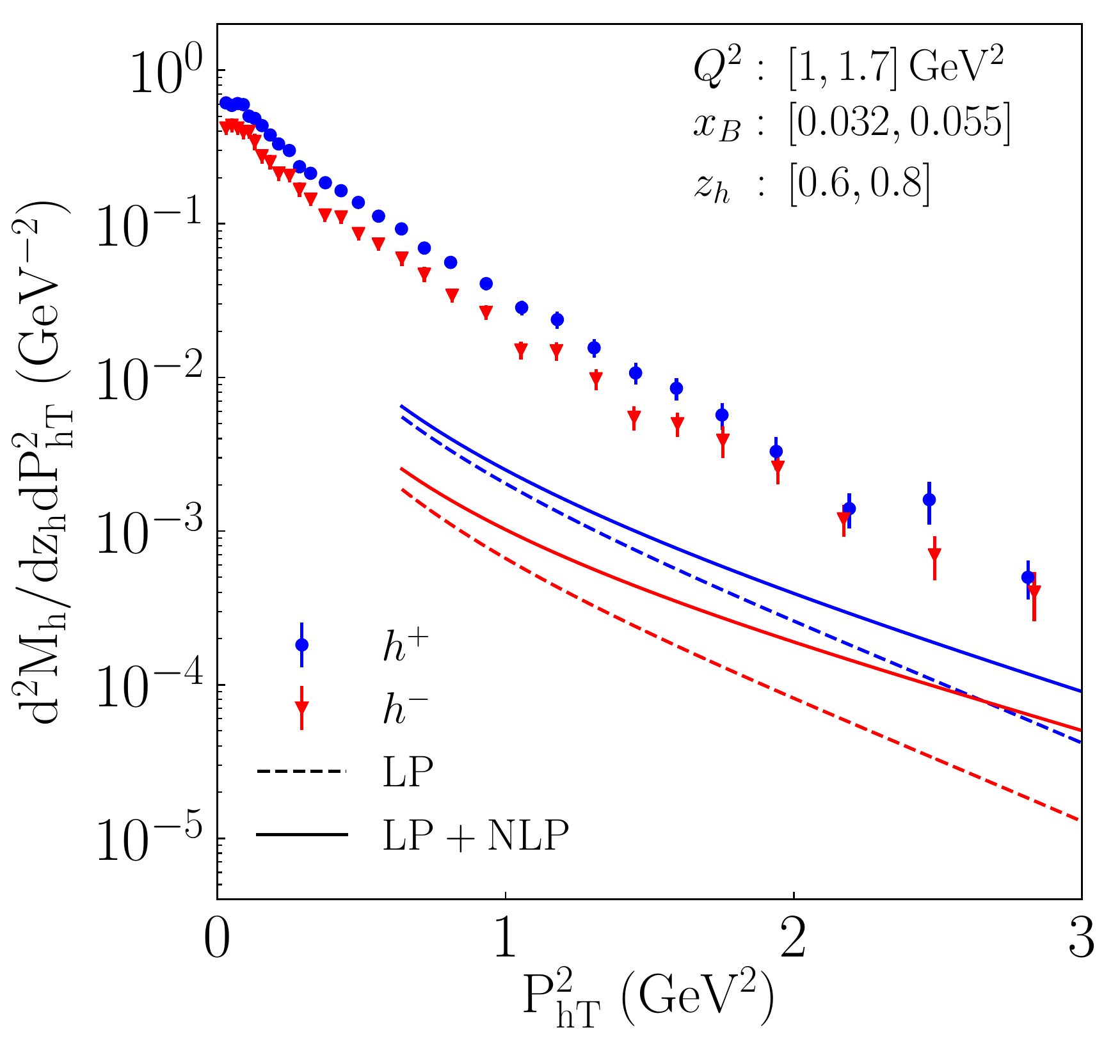}
\includegraphics[width=0.23\textwidth]{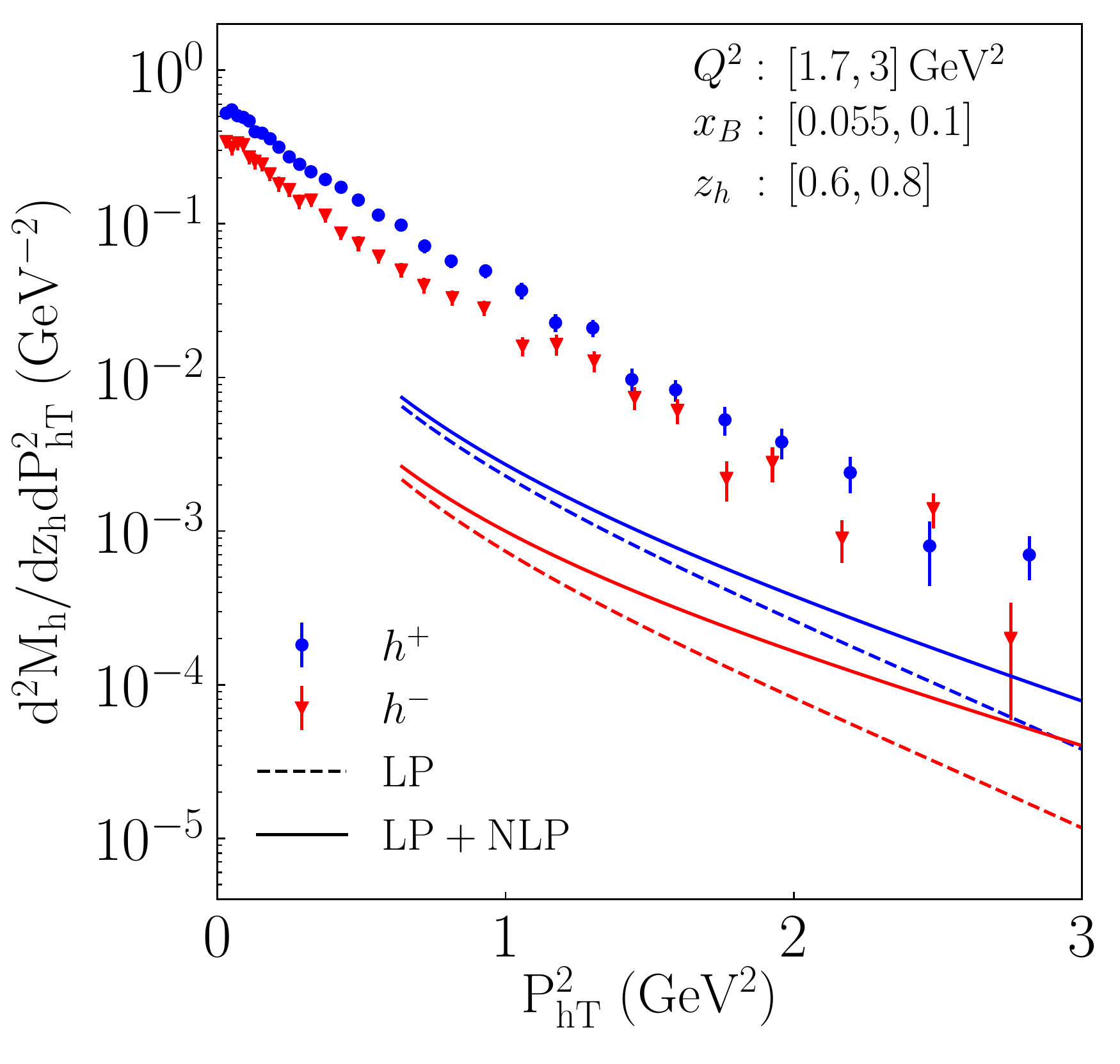}
\caption{Comparison of COMPASS data~\cite{Aghasyan:2017ctw} on the differential multiplicity with LO contribution from both LP and NLP contributions.}
\label{fig:compass-data}
\end{figure}

Since the purpose of this paper is to show the relevance and potential impact of NLP contributions to warrant a urgent and more detailed study of the NLP power corrections to the low multiplicity observables in SIDIS, instead of a precise fitting to the data, we perform straightforward leading order calculations in power of $\alpha_s$ for both LP and NLP contributions to the differential multiplicity defined in Eq.~\eqref{eq:multiplicity}.  Since the inclusive DIS cross section in the denominator is dominated by the low $P_{h_T}$ region and consistent with the LP contribution alone, we include both LP and NLP contributions to the SIDIS cross section in the numerator while having the LP contribution to the inclusive DIS cross section in the denominator in our numerical evaluation of the differential multiplicity below.  For PDFs, we use CT14 PDF set in our numerical evaluation~\cite{Dulat:2015mca}. For single-parton FFs in the LP contribution, we use the NNFF1.0 FF sets~\cite{Bertone:2017tyb}. For the NLP contribution, we only consider the channels with produced quark-antiquark pair matching the valence flavors in the color singlet and axial-vector spin projection. In addition, we approximate, $|h(P_h)X\rangle \approx |h(P_h)\rangle$, in the definition of quark-antiquark FFs, as discussed in Sec.~\ref{sec:nlp-fragmentation}, and express the quark-antiquark FFs in terms of two distribution amplitudes, as in Eq.~\eqref{eq:qqbarFFs}.  For the distribution amplitude, we adopt those from 
Ref.~\cite{Chang:2013pq}. Our numerical results are shown in Fig.~\ref{fig:compass-data} along with COMPASS data~\cite{Aghasyan:2017ctw}.
Consistent with what was found in Refs.~\cite{Gonzalez-Hernandez:2018ipj,Wang:2019bvb}, the LP contribution alone (the dotted curves) is about one order of magnitude smaller than the data.  While adding next-to-leading order corrections to the LP contribution does not help much~\cite{Wang:2019bvb}, it is clear from Fig.~\ref{fig:compass-data} that the NLP contribution (the difference between the solid and dotted curves) is large, and could be as large as a factor of five of the LP contribution when $z_h$ and $P_{h_T}$ are large, near the edge of phase space.  Therefore, in this regime where the multiplicity is low and there is not much phase space for radiation (into light hadrons), it is very important to include the NLP corrections in the QCD global fitting for extracting PDFs and FFs.  It is also an opportunity for studying QCD power corrections and the formation or emergence of hadrons from perturbatively produced quarks and gluons.

\section{Discussions and future opportunities \label{sec:discussion}}

As emphasized earlier, it is not our goal of this paper to fit the COMPASS data to extract the NLP contributions, since the precise LP and NLP contributions to one physical observable, or more precisely, to the differential multiplicity in Eq.~(\ref{eq:multiplicity}), depend on more than one unknown, nonperturbative function.  In principle, we need theoretical calculations for more physical observables, which are also sensitive to the same quark-antiquark FFs, and corresponding data to perform QCD global analyses to extract both PDFs and FFs, as well as these new quark-antiquark FFs, which could provide much more insights to the color neutralization and formation of light hadrons, complimentary to what we have learned from the LP single-parton FFs.  The predictive power of this QCD factorization approach beyond the LP is our ability to calculate the short-distance hard parts and the universality of these new multiparton FFs.  

In this section, we will discuss the source of possible contributions to these new quark-antiquark FFs to gain some insights into their potential structure and functional forms, and to identify new physical observables that could also be sensitive to the same quark-antiquark FFs, so that we could test the universality of these new multiparton FFs and QCD dynamics beyond the LP contributions in the future work.
\begin{figure}[htp]
\centering
$\vcenter{\hbox{\includegraphics[width=0.19\textwidth]{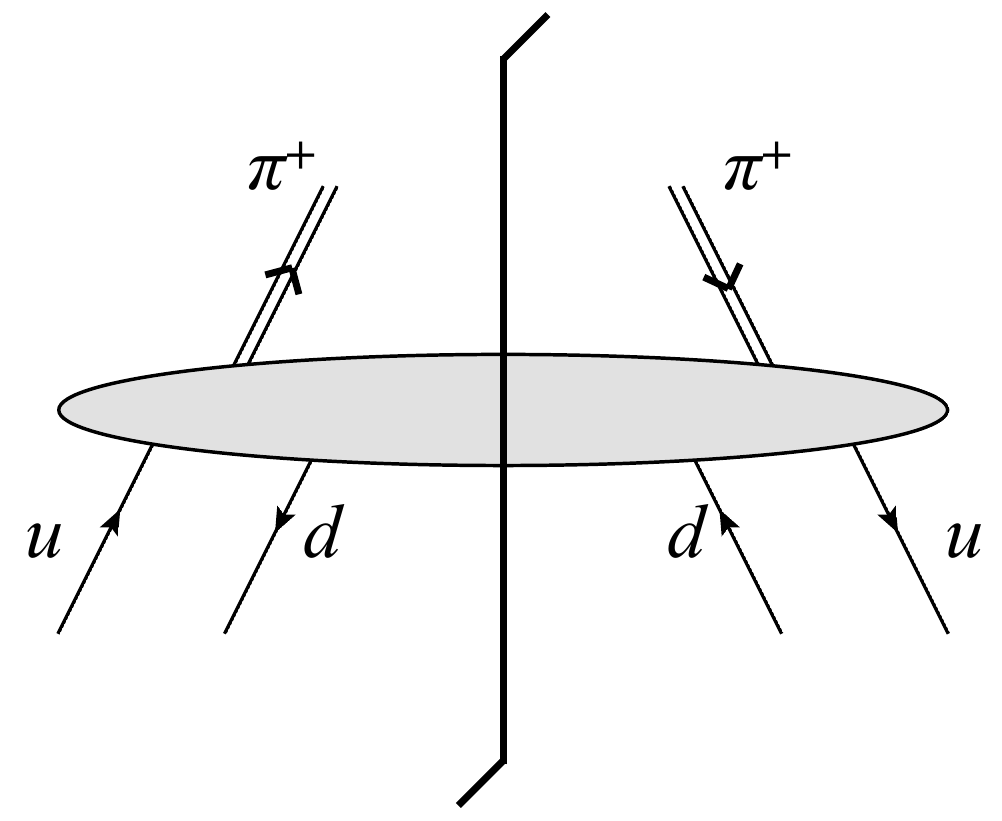}}}
\approx
\vcenter{\hbox{\includegraphics[width=0.19\textwidth]{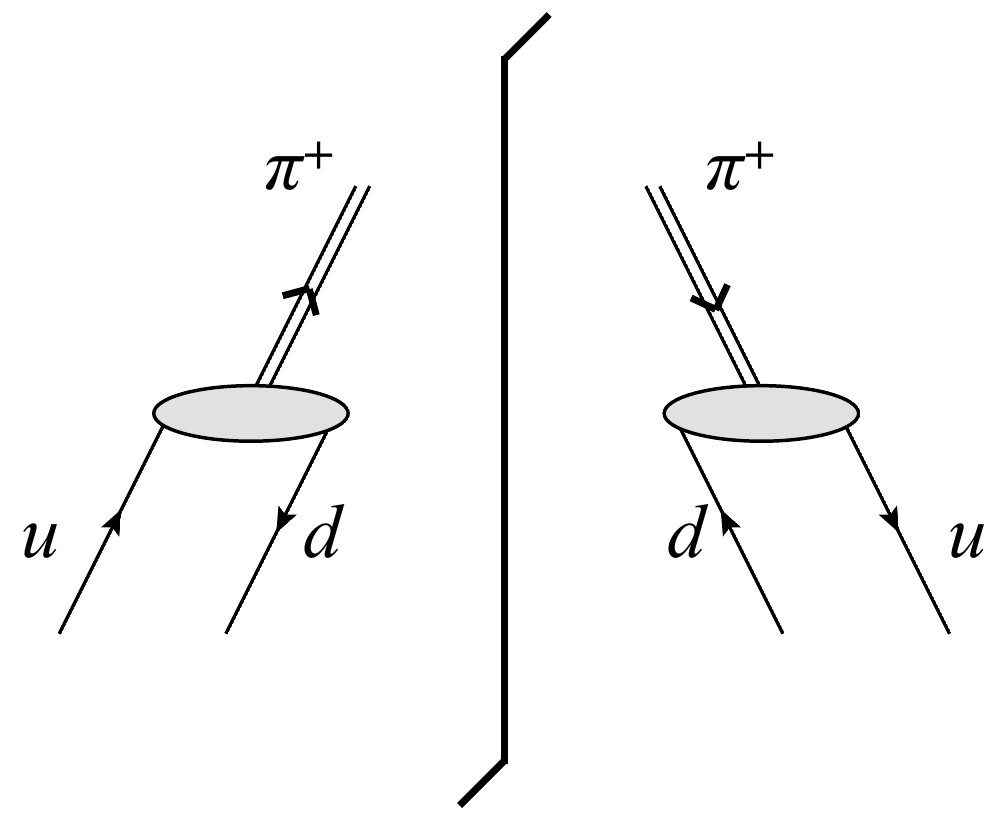}}} 
\quad $
\\
$+
\vcenter{\hbox{\includegraphics[width=0.19\textwidth]{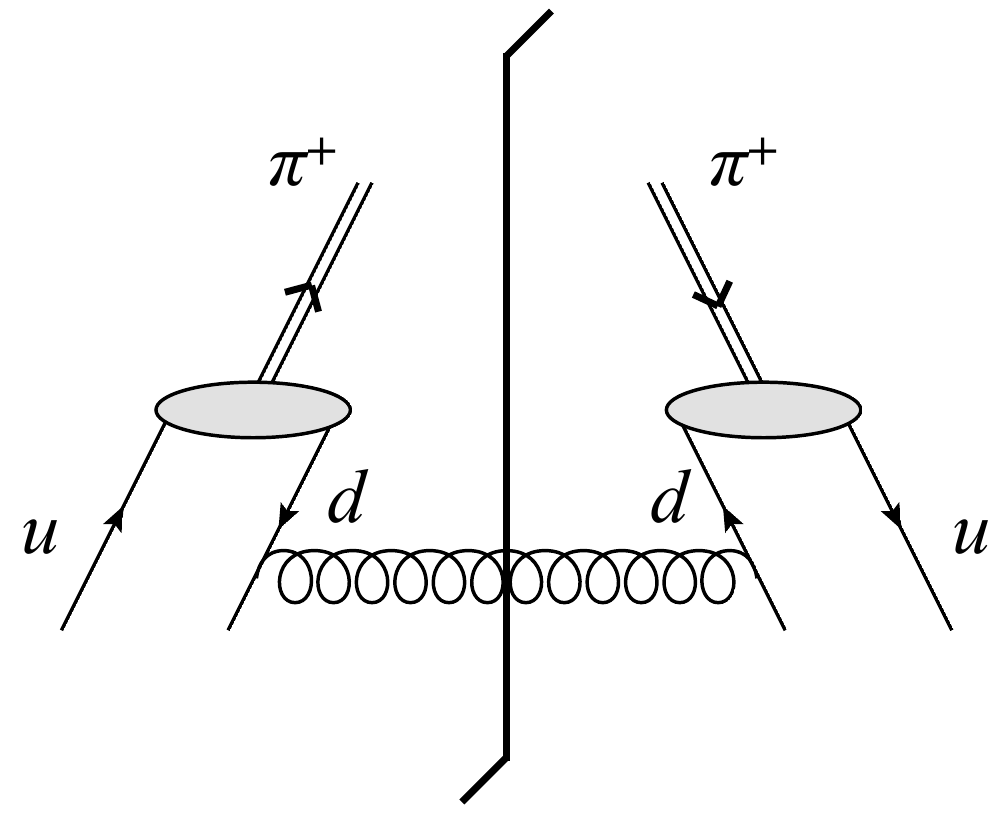}}}
+
\vcenter{\hbox{\includegraphics[width=0.19\textwidth]{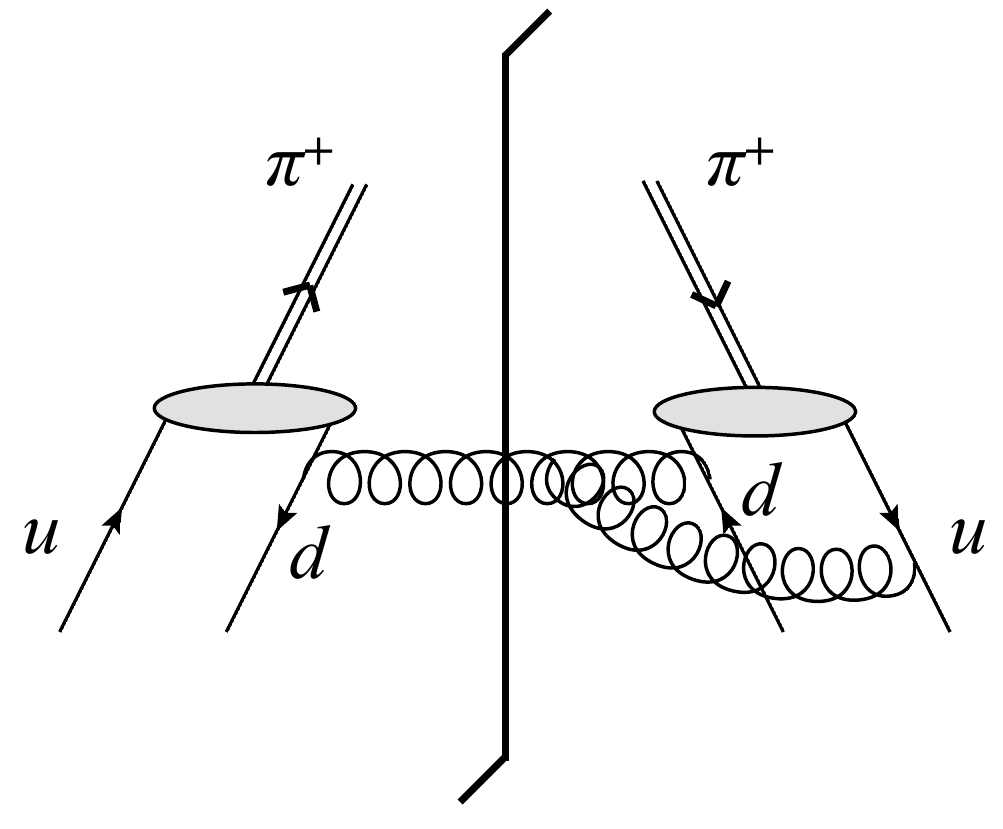}}}
+\cdots$
\caption{Feynman diagram representation of the FFs for a pair of $u\bar{d}$ to fragment into a $\pi^+$ meson.}
\label{fig:qqbarFFs}
\end{figure}

\begin{figure}[htp]
\centering
$\vcenter{\hbox{\includegraphics[width=0.19\textwidth]{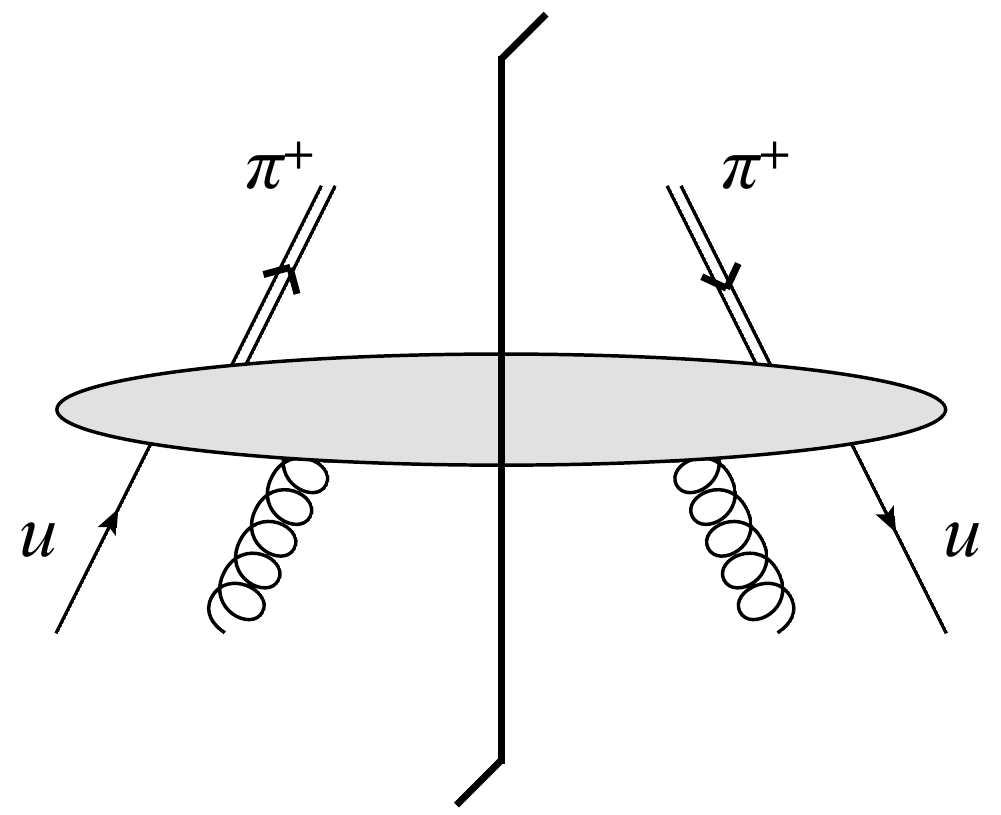}}}
\approx
\vcenter{\hbox{\includegraphics[width=0.19\textwidth]{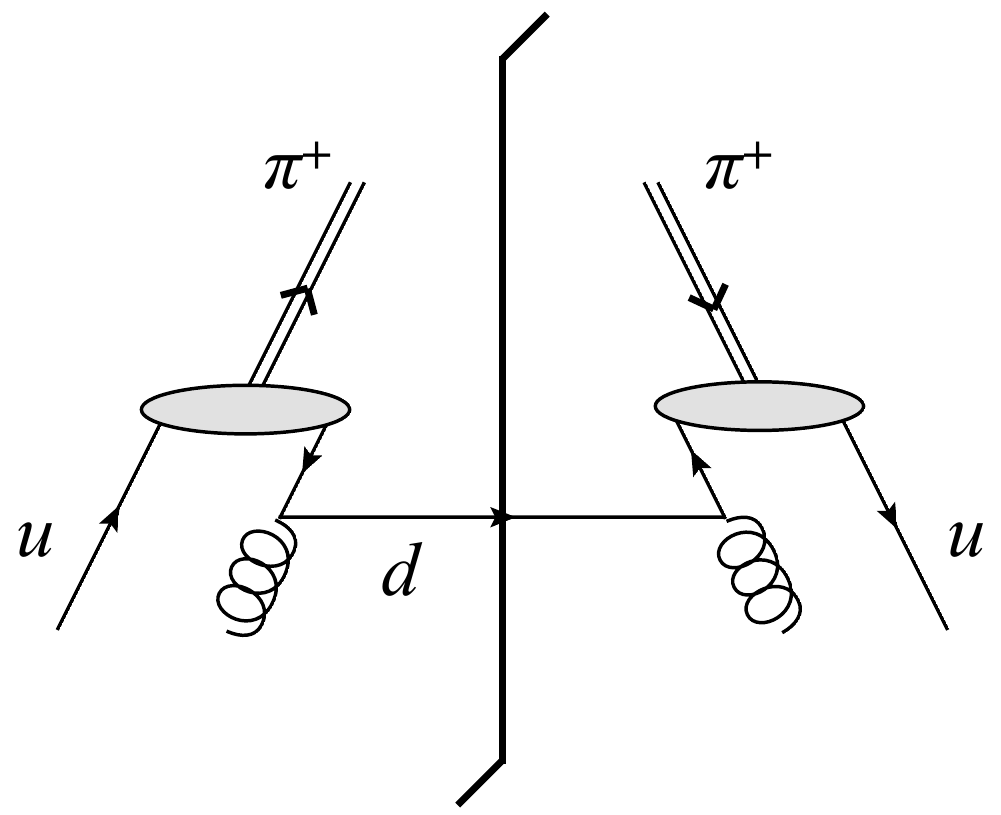}}}
+\cdots$
\caption{Feynman diagram representation of the FFs for a $ug$ pair to fragment into a $\pi^+$ meson.}
\label{fig:qgFFs}
\end{figure}

The quark-antiquark FFs are nonperturbative functions and cannot be calculated within QCD perturbation theory.  However, like PDFs and FFs, we might be able to gain some insights into these nonperturbative functions' asymptotic behavior as the variables of these functions approach to an extreme limit, such as $z\to 1$ (or $x\to 1$ or $0$ in the case of PDFs).  With the operator definition in Eq.~(\ref{eq:qq2piFF}), in principle, we could represent these quark-antiquark FFs in terms of Feynman diagrams --a Feynman diagram representation.  For example, the FFs for a $u\bar{d}$ pair to fragment into a $\pi^+$ could be represented by an infinite number of Feynman diagrams, as shown in Fig.~\ref{fig:qqbarFFs}.  The first diagram on the right of the ``$\approx$" sign is effectively the lowest order diagram in power of $\alpha_s$ in the approximation, $|h(P_h)X\rangle \approx |h(P_h)\rangle$, which led to the approximated expression of $D_{[q\bar{q}'(1a)]}(z,\xi,\zeta,\mu_0)$ in Eq.~(\ref{eq:qqbarFFs}).  With additional radiation of gluons, other diagrams in Fig.~\ref{fig:qqbarFFs} could also contribute to $D_{[q\bar{q}'(1a)]}(z,\xi,\zeta,\mu_0)$, but, cannot be proportional to $\delta(1-z)$, instead, proportional to $(1-z)^n$ as $z\to 1$.  
Although the power of $n$ is a nonperturbative number and depends on the scale at which the FFs are measured, the power $n$ should be positive that leads to a powerlike suppression to the NLP contribution from these diagrams, similar to the suppression from LP single parton FFs when $z\to 1$  as discussed earlier in this paper.  In addition, a quark-gluon pair could also fragment into a meson as illustrated by Feynman diagrams in Fig.~\ref{fig:qgFFs}, which is suppressed by the power of $1-z$ as $z\to 1$.  In general, in a confining theory, like QCD, the neutralization of color of these radiated gluons (and/or quarks) requires them to turn into physical hadrons in the final-state, such as the pions with the lightest mass, which strongly suppresses their contributions to the physical cross sections near the edge of phase space (or those with a very small multiplicity).  

\begin{figure}[htp]
\centering
$\vcenter{\hbox{\includegraphics[width=0.19\textwidth]{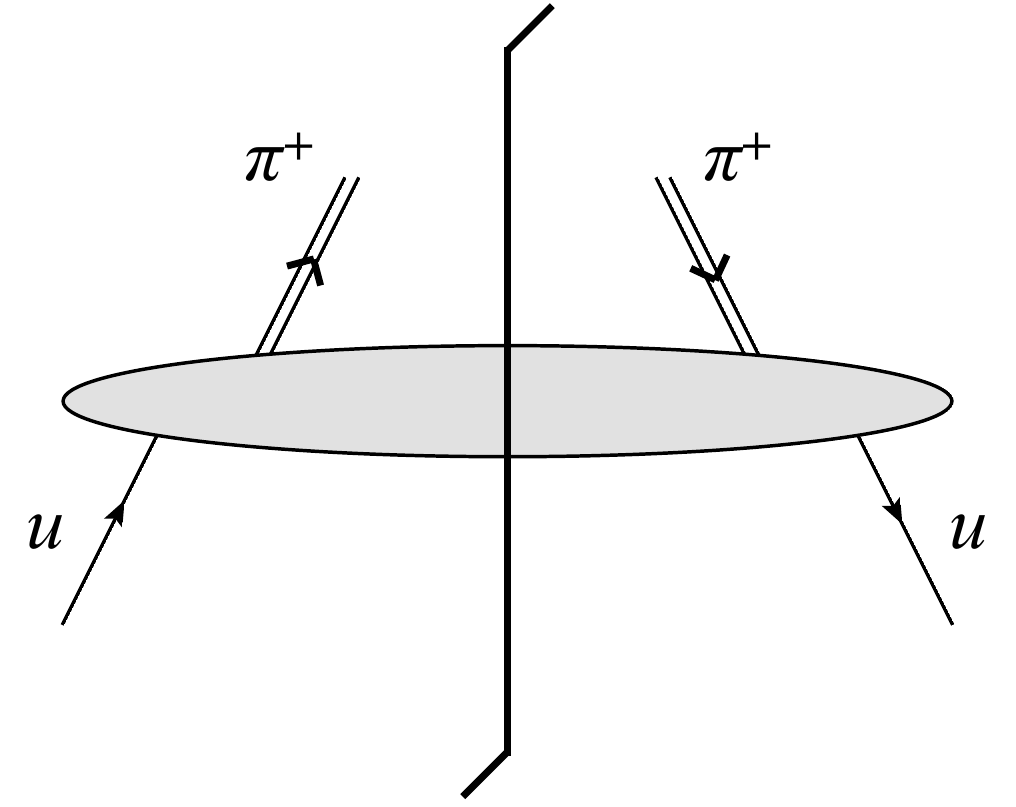}}}
\approx
\vcenter{\hbox{\includegraphics[width=0.19\textwidth]{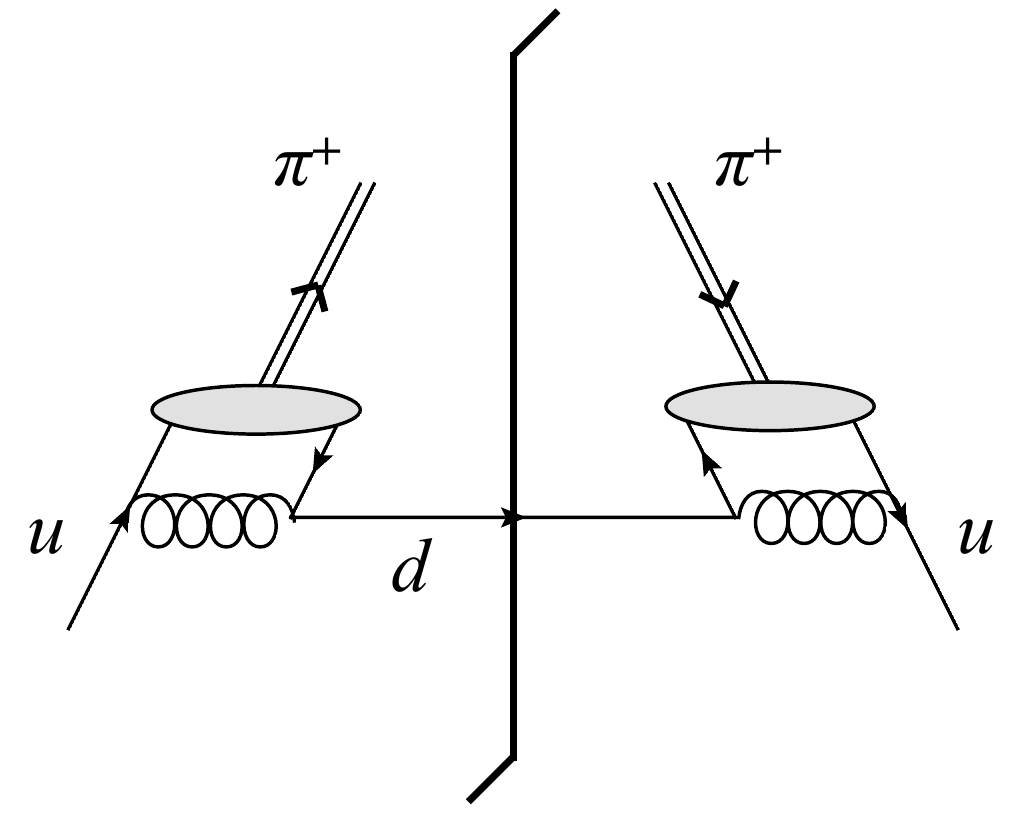}}}
+\cdots$
\caption{Feynman diagram representation of the single-parton FFs to a $\pi^+$ meson via an intermediate $u\bar{d}$ pair.}
\label{fig:mix}
\end{figure}

Although the factorized NLP contribution to the SIDIS cross sections in Eq.~(\ref{eq:factorize}) is formally suppressed by the hard scale $P_{h_T}$, the impact of the NLP contribution to the physical cross sections does not vanish as the power of $1/P_{h_T}$ \cite{Kang:2014tta,Mueller:1985wy,Qiu:1986wh}.  With the factorization formula in Eq.~(\ref{eq:factorize}), which is a factorization of perturbative {\it collinear} singularities of partonic scattering, we must modify the Dokshitzer-Gribov-Lipatov-Altarelli-Parisi (DGLAP) evolution of LP single-parton FFs to be consistent to the {\it collinear} factorization accuracy at the NLP.  Following the discussion in Ref.~\cite{Kang:2014tta}, we can derive the evolution equations for both the single-parton and double-parton FFs from the factorization formalism in Eq.~(\ref{eq:factorize}).  Since a physical observable should be independent of the choice of the factorization scale, we have 
\begin{align}
& \frac{d}{d \ln\mu^2}\left( D_{f\to h}\otimes d\hat{\sigma}_{\gamma^{(*)}+A\to f+X} \right.
\nonumber\\
& {\hskip 0.35in} \left.
+ D_{[ff'(\kappa)]\to h} \otimes d\hat{\sigma}_{\gamma^{(*)}+A\to [ff'(\kappa)]+X}
\right)=0\, ,
\label{eq:invariant}
\end{align}
where $\otimes$ represents the convolution of momentum fractions as defined in Eq.~(\ref{eq:factorize}). From Eq.~(\ref{eq:invariant}), we obtain a closed set of evolution equations for the FFs~\cite{Kang:2014tta}
\begin{align}
&\frac{\partial}{\partial \ln\mu^2} D_{[ff'(\kappa)]\to h}
\nonumber \\
& {\hskip 0.3in} =
\sum_{[ff'(\kappa')]} 
D_{[ff'(\kappa')]\to h}\otimes \Gamma_{[ff'(\kappa')]\to [ff'(\kappa')]}\, ;
\label{eq:NLP}
\end{align}
and
\begin{align}
&\frac{\partial}{\partial \ln\mu^2} D_{f\to h}
=
\sum_{f'} D_{f\to h}\otimes \gamma_{f\to f'}
\nonumber \\
&{\hskip 0.3in}+ 
\frac{1}{\mu^2} 
\sum_{[ff'(\kappa')]} 
D_{[ff'(\kappa')]\to h}\otimes \widetilde{\gamma}_{f\to [ff'(\kappa')]} \, ,
\label{eq:LP}
\end{align}
where $\Gamma_{[ff(\kappa')]\to [ff'(\kappa')]}$ is the evolution kernel for resumming logarithmic collinear contribution to the double-parton FFs, $\gamma_{f\to f'}$ is the normal LP DGLAP-type evolution kernel for resumming logarithmic collinear contribution to the one-parton FFs, and $\widetilde{\gamma}_{f\to [ff'(\kappa')]}$ is a new-type of evolution kernel for resumming the {\it collinear} contributions from the diagrams such as those in Fig.~\ref{fig:mix} to the one-parton FFs.  
Although the second terms on the right-hand side of Eq.~(\ref{eq:LP}) is power suppressed by $1/\mu^2$, its contribution to the physical observables, such as the SIDIS cross section in Eq.~(\ref{eq:factorize}), does not vanish as $1/P_{h_T}$ since it contributes to the {\it slope} of $D_{f\to h}$, not the $D_{f\to h}$ itself \cite{Kang:2014tta,Mueller:1985wy,Qiu:1986wh}.  That is, in order to understand the true impact of the NLP contribution to the SIDIS, we need to do a simultaneous QCD global fitting of PDFs and FFs~\cite{Sato:2019yez}, together with double-parton FFs if one wants to include the COMPASS data or other data near the edge of phase space.  

To avoid the difficulty of having too many double-parton FFs, as the leading approximation, it might be practical for now to make the ``valence quark'' approximation to keep the quark-antiquark flavors --$[ff'(\kappa)]$ in both the cross section calculations and evolution equations to be the same as the valence quark flavors of the observed meson.  That is, for the double-parton fragmentation functions,  we ignore the contributions from the double-parton FFs in Fig.~\ref{fig:qgFFs}, while keeping the contributions from the diagrams in Fig.~\ref{fig:qqbarFFs}.

To close this section, we estimate the NLP contribution to charged pion and kaon productions in upcoming SIDIS experiments at Jefferson Lab (JLab), which have much lower collision energies than what COMPASS had, and thus, should have much less high multiplicity events. Therefore, the impact of the NLP contribution at JLab kinematics could be more significant than that at COMPASS kinematics. 
\begin{figure}[htp]
\centering
\includegraphics[width=0.235\textwidth]{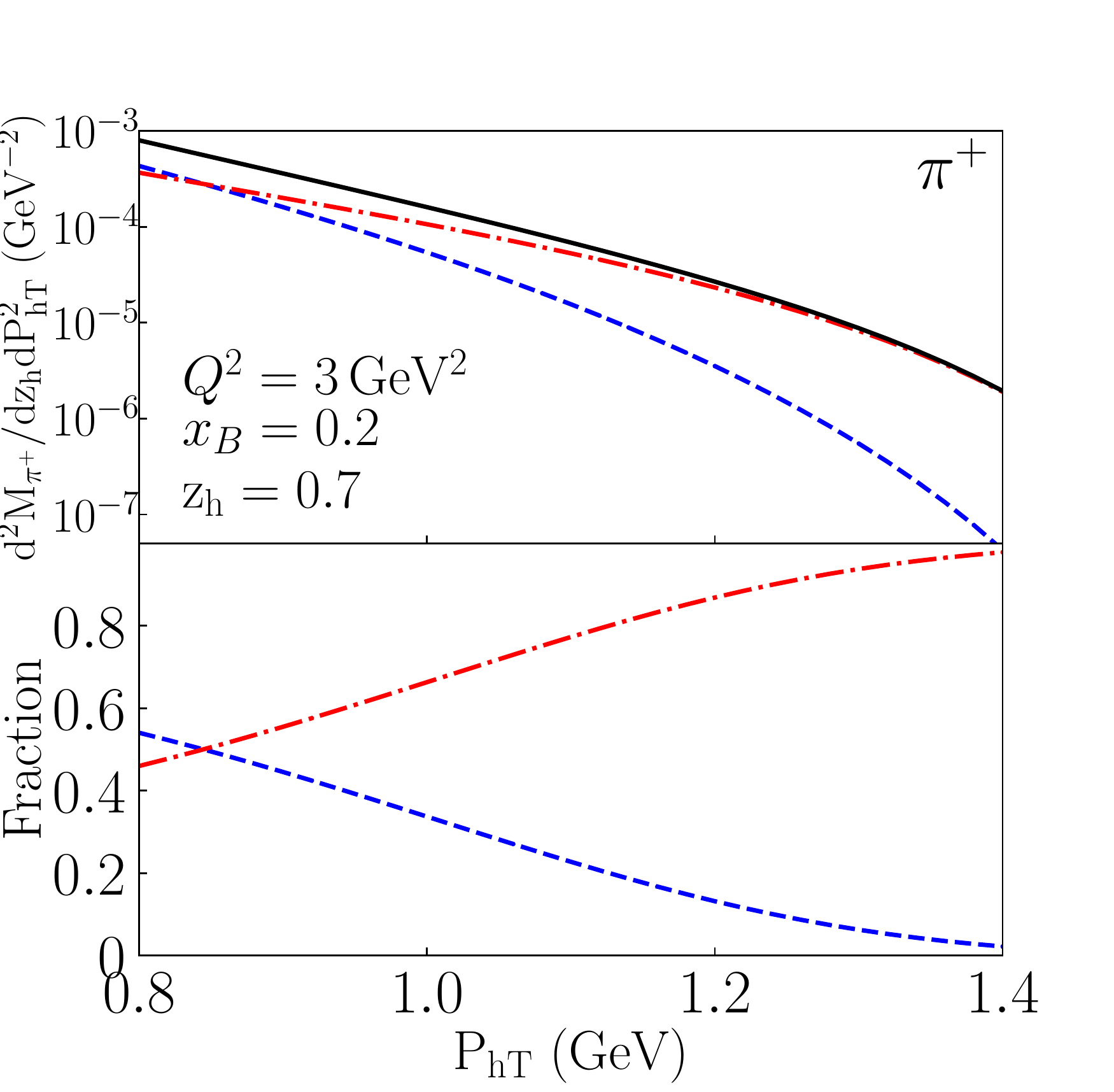}
\includegraphics[width=0.235\textwidth]{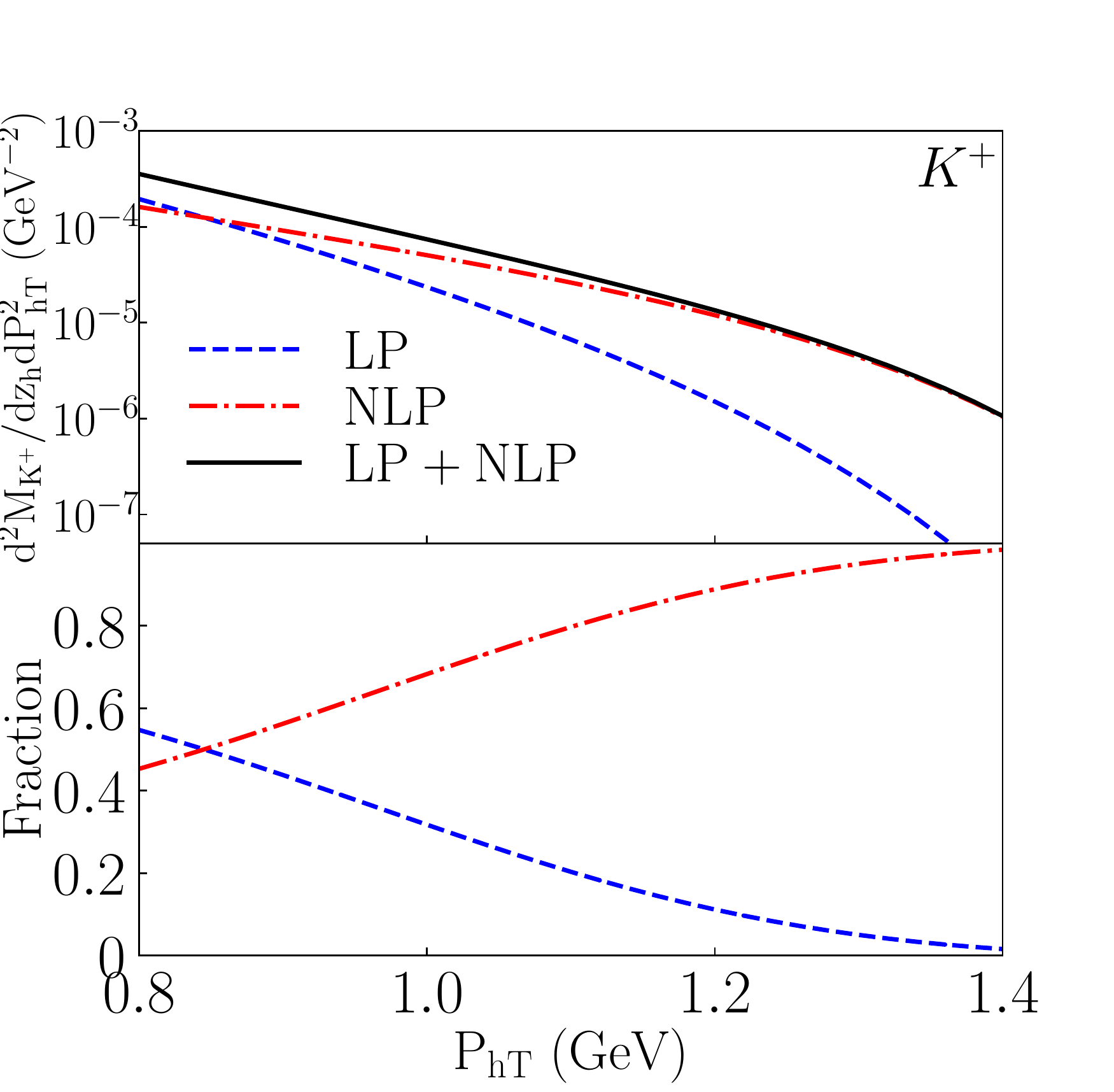}
\caption{Differential multiplicities at JLab kinematics: $E_{\rm beam}=11\,\rm GeV$, $Q^2=3\,\rm GeV^2$, $x_B=0.2$, and $z_h=0.7$.}
\label{fig:jlab12pt}
\end{figure}
In Fig.~\ref{fig:jlab12pt}, we present our calculated  differential multiplicities, defined in Eq.~(\ref{eq:multiplicity}), for both pion and kaon production in SIDIS experiments with a typical kinematics at JLab.  For the kaon distribution amplitude, we adopt those in Ref.~\cite{Shi:2014uwa}. We show the LP (dashed), NLP (dot-dashed) and LP+NLP (solid) contributions, respectively. As expected, the NLP term dominates large-$P_{h_T}$ region, due to the strong $(1-z)$-power suppression from the single-parton FFs to the LP contribution, even though the NLP contribution is formally suppressed by extra power of $1/P_{h_T}$.  To make this point even more quantitative and transparent, we plot the fractional contribution to the differential multiplicity from the LP (dashed) and NLP (dot-dashed) in the lower panels, respectively.  

\begin{figure}[htp]
\centering
\includegraphics[width=0.235\textwidth]{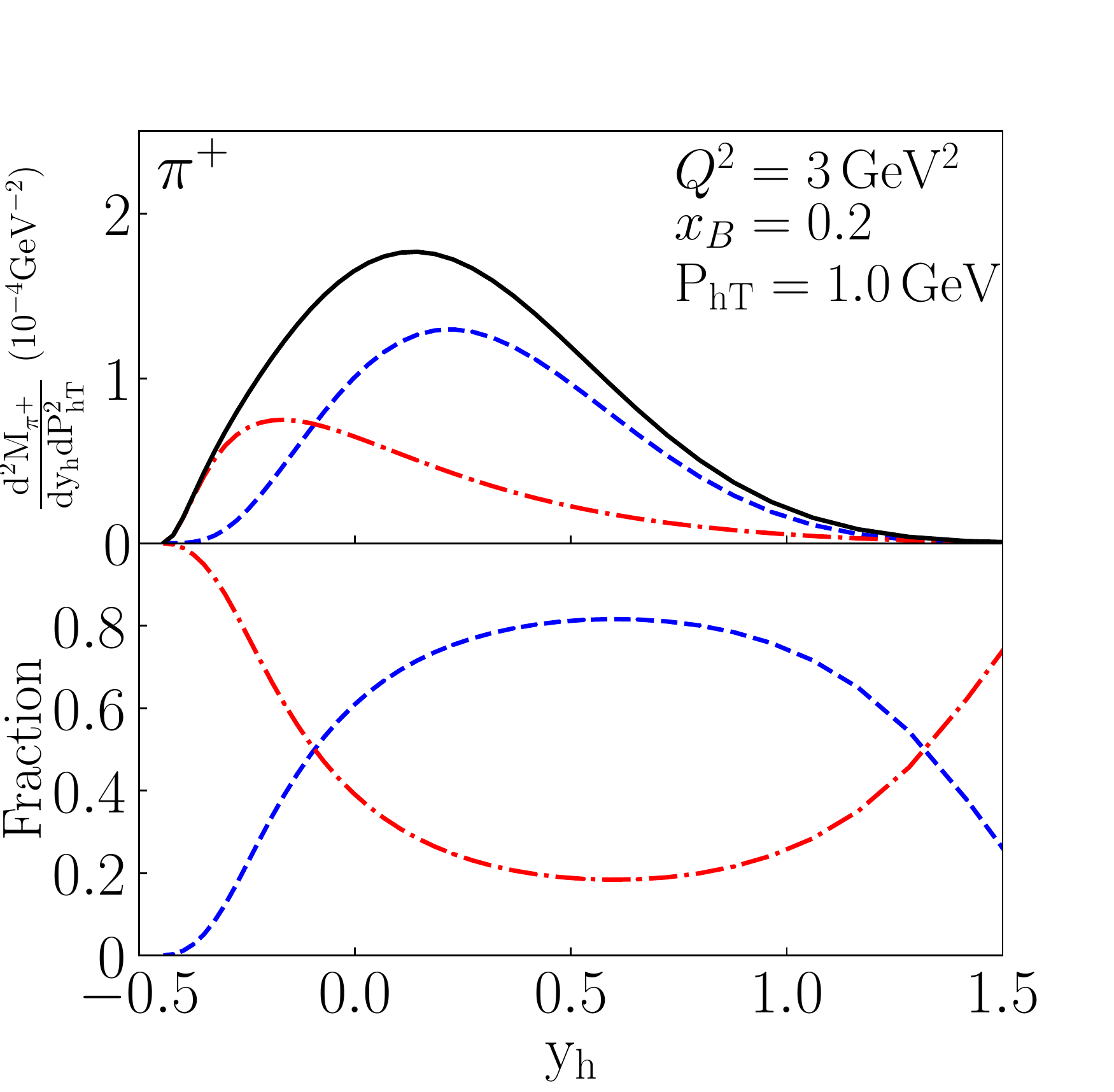}
\includegraphics[width=0.235\textwidth]{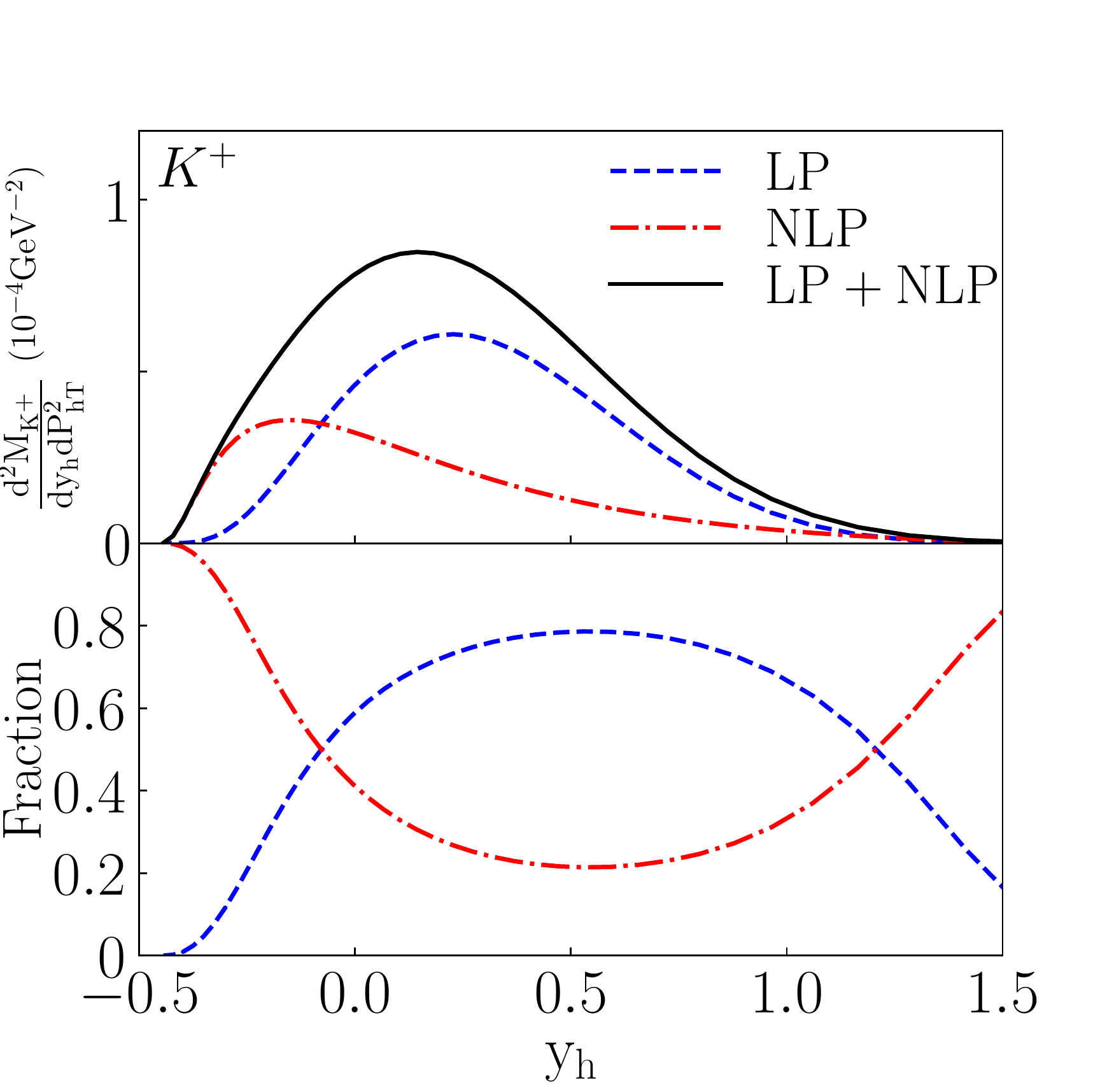}
\caption{Rapidity distributions of LP and NLP contributions at JLab kinematics: $E_{\rm beam}=11\,\rm GeV$, $Q^2=3\,\rm GeV^2$, $x_B=0.2$, and $P_{h_T}=1.0\,\rm GeV$.}
\label{fig:jlab12yh}
\end{figure}
In addition to the $P_{h_T}$ dependence, the NLP contribution has different rapidity distributions from that of the LP contribution. In Fig.~\ref{fig:jlab12yh}, we show the LP, NLP and LP+NLP contributions separately as a function of the rapidity $y_h$ of the measured charged meson. The rapidity $y_h$ is defined in the photon-target collinear frame with virtual photon momentum $q=(-Q/\sqrt{2},Q/\sqrt{2},{\bf 0}_\perp)$. As expected, the NLP contribution favors more negative rapidity, which corresponds to larger $z_h$ regime where the LP contribution is suppressed by powers of $(1-z)$ from its single-parton FFs. Because the phase space for the produced partons to radiate to light hadrons is smaller at JLab energies, the NLP contribution is sizable, about 20\%, even in the midrapidity region comparing with the LP contribution. Therefore, more detailed studies of the NLP corrections  are urgent for upcoming SIDIS experiments at JLab.

\section{Summary and conclusions \label{sec:summary}}

We have presented the first calculations of power corrections to charged meson productions in SIDIS at large transverse momentum, $P_{h_T}\sim Q\gg \Lambda_{\rm QCD}$, in terms of the QCD collinear factorization formalism. We found that the power corrections are very important for the events near the edge of phase space where the hadron multiplicity is low, and $z_h\to 1$ and $P_{h_T}$ is large. 

By expanding the SIDIS cross section in terms of the inverse power of the observed large momentum scale, $1/P_{h_T}$ or $1/Q$, we perturbatively factorized the leading and the first subleading power contributions into short-distance hard parts and corresponding nonperturbative PDFs and FFs. The first subleading power (or the NLP) contributions are formally suppressed by powers of the large momentum scale in the hard part comparing to the LP contributions, but, as we have shown in this paper, the net size of its contribution to the cross section is not necessarily smaller than the formal LP contribution, since the hadronization for the produced quark-antiquark pair might be more effective when the pair has the right quantum number to match to the measured meson. We demonstrated that contributions from such more direct production channels are very important for cross sections near the edge of phase space where events are typically with a low multiplicity.  When the phase space for parton shower is closing out at the edge of the kinematic limit, it makes much harder for the fragmenting quark of the LP contribution to neutralize its color, which is consistent with the $(1-z)$-type power suppression from the single-parton FFs when $z\to 1$. On the other hand, for the NLP contribution, the leading transition for the produced quark-antiquark pair to a measured meson could be proportional to $\delta(1-z)$ without the phase space suppression, which could help make up the $1/P_{h_T}$-type suppression to produce the quark-antiquark pair at the NLP instead of a single fragmenting parton at the LP.

To quantitatively estimate the size of the NLP contributions, possibly its lower limit, we performed a calculation considering only the partonic subprocesses that produce a quark-antiquark pair with the same flavor combination of the valence constituents of the detected meson, because these subprocesses could have the best chance to compete with LP channels. From the operator definition of the quark-antiquark FFs, and an approximation for the hadronic final-state of the FFs, $|h(P_h)X\rangle\approx |h(P_h)\rangle$, we were able to express the unknown quark-antiquark FFs to a charged meson in terms of this meson's distribution amplitudes, which are better studied.  With the perturbatively calculated LO hard parts and the approximated quark-antiquark FFs, we were able to present numerically the approximate size of the NLP contributions, and in particular, to demonstrate quantitatively how important the NLP contributions are in comparison with the LP contributions for the COMPASS kinematics, as well as the energy regime at JLab.  At large $z_h$ and $P_{h_T}$, we found that the NLP contribution could be as large as five times of the LP contribution. Although the purpose of this paper is not to fit COMPASS data, our finding ensures the importance to take into account NLP corrections in the QCD global analysis if the COMPASS data or other data near the edge of phase space are going to be included in the analyses. Therefore, it is urgent to have more detailed studies on the NLP corrections.

The new double-parton FFs could provide more insights to the color neutralization and the formation of light hadrons in complimentary to the knowledge we have learned from the single-parton FFs. As nonperturbative functions, they cannot be calculated within the QCD perturbation theory, and a simultaneous QCD global fitting of PDFs, single-parton FFs, and double-parton FFs, together with  more physical observables that are sensitive to double-parton FFs is needed. The DGLAP evolution equation of the single-parton FFs should also be modified consistently to the accuracy at the NLP. Though the correction term from the NLP is suppressed by $1/\mu^2$ in the evolution equation, its effect on physical observables does not vanish even at high scales, because it contributes to the slope of FFs \cite{Kang:2014tta,Mueller:1985wy,Qiu:1986wh}. The predictive power of such factorization approach is the universality of the new FFs, together with our ability to calculate the short-distance hard parts. Upcoming experiments at JLab, as well as those at the future EIC, will provide ample  opportunities to study the multiparton FFs to investigate a new domain of QCD dynamics sensitive to multiparton correlations.

\acknowledgments{
We would like to thank J.-P.~Chen, Z.-t.~Liang, T.C.~Rogers, N.~Sato, Y.-K.~Song, and G.~Sterman for helpful discussions. This work is supported in part by the U.S. Department of Energy, Office of Science, Office of Nuclear Physics under contract No. DE-AC05-06OR23177, within the framework of the TMD Topical Collaboration, and No. DE-FG02-03ER41231.
}


\end{document}